\documentclass[prd,twocolumn,superscriptaddress,floatfix,amsmath,amssymb,amsfonts,nofootinbib,longbibliography]{revtex4-2}

\usepackage{float} 
\usepackage{scalerel}
\usepackage[normalem]{ulem}
\usepackage[english]{babel}
\usepackage{graphicx}
\usepackage{dcolumn}
\usepackage{bm}
\usepackage{blindtext}
\usepackage{verbatim}
\usepackage{relsize}
\usepackage{mathrsfs}
\usepackage{musicography}
\usepackage{amsmath}
\usepackage{blindtext}
\usepackage{cancel}
\usepackage{physics}
\usepackage{epstopdf}
\usepackage{mathtools}
\usepackage{blindtext}
\usepackage{tensor}
\usepackage{color}
\usepackage[usenames,dvipsnames]{pstricks}
\usepackage{epsfig}
\usepackage{pst-grad} 
\usepackage{pst-plot} 
\usepackage{hyperref}
\usepackage{verbatim}
\usepackage{slashed}
\usepackage{dsfont}

\newcommand{\mf}{\mathsf}

\newcommand{\ii}{\mathrm{i}}

\renewcommand{\r}{\hat{\rho}}

\newcommand{\tc}[1]{\textsc{#1}}
\newcommand{\trr}[1]{\textcolor{red}{#1}}
\newcommand{\tbb}[1]{\textcolor{blue}{#1}}

\newcommand{\N}{{{}_{\!\bm N\!}}}

\makeatletter 

\renewcommand\onecolumngrid{
\do@columngrid{one}{\@ne}%
\def\set@footnotewidth{\onecolumngrid}
\def\footnoterule{\kern-6pt\hrule width 1.5in\kern6pt}%
}

\renewcommand\twocolumngrid{
        \def\footnoterule{
        \dimen@\skip\footins\divide\dimen@\thr@@
        \kern-\dimen@\hrule width.5in\kern\dimen@}
        \do@columngrid{mlt}{\tw@}
}%
\makeatother

\allowdisplaybreaks[1] 

\begin{document}

\title{Fully Relativistic Entanglement Harvesting}

\author{T. Rick Perche}
\email{trickperche@perimeterinstitute.ca}

\affiliation{Perimeter Institute for Theoretical Physics, Waterloo, Ontario, N2L 2Y5, Canada}
\affiliation{Department of Applied Mathematics, University of Waterloo, Waterloo, Ontario, N2L 3G1, Canada}
\affiliation{Institute for Quantum Computing, University of Waterloo, Waterloo, Ontario, N2L 3G1, Canada}

\author{Jos\'e Polo-G\'omez}
\email{jpologomez@uwaterloo.ca}

\affiliation{Perimeter Institute for Theoretical Physics, Waterloo, Ontario, N2L 2Y5, Canada}
\affiliation{Department of Applied Mathematics, University of Waterloo, Waterloo, Ontario, N2L 3G1, Canada}
\affiliation{Institute for Quantum Computing, University of Waterloo, Waterloo, Ontario, N2L 3G1, Canada}

\author{Bruno de S. L. Torres}
\email{bdesouzaleaotorres@perimeterinstitute.ca}\affiliation{Perimeter Institute for Theoretical Physics, Waterloo, Ontario, N2L 2Y5, Canada}
\affiliation{Institute for Quantum Computing, University of Waterloo, Waterloo, Ontario, N2L 3G1, Canada}
\affiliation{Department of Physics and Astronomy, University of Waterloo, Waterloo, Ontario, N2L 3G1, Canada}

\author{Eduardo Mart\'in-Mart\'inez}
\email{emartinmartinez@uwaterloo.ca}

\affiliation{Perimeter Institute for Theoretical Physics, Waterloo, Ontario, N2L 2Y5, Canada}
\affiliation{Department of Applied Mathematics, University of Waterloo, Waterloo, Ontario, N2L 3G1, Canada}
\affiliation{Institute for Quantum Computing, University of Waterloo, Waterloo, Ontario, N2L 3G1, Canada}

\begin{abstract}
    We study the protocol of entanglement harvesting when the particle detectors that harvest entanglement from the field are replaced by fully relativistic quantum field theories. We show that two localized modes of the quantum field theories are able to harvest the same amount of leading order entanglement as two non-relativistic particle detectors, thus implying that QFT probes can generally harvest more entanglement than particle detectors. These results legitimize the use of particle detectors to study entanglement harvesting regardless of their internally non-relativistic nature.

\end{abstract}

\maketitle

\section{Introduction}

    Entanglement is a fundamental feature of quantum theories whose measurement cannot be reproduced by local classical models. Its applications range from improving quantum communication protocols~\cite{EntanglementBasedQC}, cryptography~\cite{Ekert1991} and facilitating computational tasks~\cite{Schor}. In this sense, entanglement is a quantum resource that can be used to enhance our computational power. Being at the heart of quantum theory, one could expect that entanglement is well understood in most relevant scenarios, and indeed, there are situations where a full characterization of entanglement is known, such as for example for arbitrary states in sufficiently simple bipartite systems. However, quantifying entanglement in mixed states of arbitrary systems~\cite{HorodeckiReview} and in infinite dimensional Hilbert spaces is still an ongoing research topic~\cite{entInfiniteDim}. 
    
    The situation is even worse in the case of quantum field theory (QFT), where, strictly speaking, the Hilbert space cannot be factorized as a tensor product of local Hilbert spaces associated to causally disjoint regions. Therefore, even the well-established measures of entanglement for bipartite pure states in quantum mechanics are not very meaningful in QFT. 
    One of the main reasons for this is because any sufficiently regular QFT state restricted to two non-complementary regions separated by some finite distance will be mixed. As a consequence, the entanglement shared between the two subsystems of interest is very hard to characterize---both because of our limited understanding of mixed-state entanglement in general, and because local subregions in QFT are associated to type III von Neumann algebras, and therefore do not even admit descriptions in terms of density matrices~\cite{Witten, vNalgebrasSorceReview}. In this light, other techniques for studying the entanglement between subregions of a QFT have been developed. One of them, on which we will focus in this work, is the protocol of entanglement harvesting.
    
    Entanglement harvesting is a protocol in which localized quantum probes couple to a quantum field aiming to extract entanglement from it~\cite{Valentini1991,Reznik2003,Pozas-Kerstjens:2015}. If the probes are unable to communicate through the field\footnote{By `cannot communicate' we mean that the detectors cannot exchange information via the propagation of their respective actions on the field (e.g., when they are spacelike separated). If the detectors can communicate through the field, the acquired entanglement may not originate from extracting preexisting field correlations. For a discussion on this, see~\cite{ericksonNew}.}, the entanglement they acquired serves as a witness for the entanglement in the field between the regions that they couple to~\cite{kelly}. This simple approach to quantifying entanglement in a quantum field theory has the advantages of being readily applicable for quantum fields in both flat~\cite{Valentini1991,Reznik2003,Reznik1,reznik2,Salton:2014jaa,Pozas-Kerstjens:2015,Pozas2016,HarvestingSuperposed,Henderson2019,bandlimitedHarv2020,ampEntBH2020,carol,boris,ericksonNew,threeHarvesting2022,twist2022,cisco2023harvesting} and curved spacetimes~\cite{Ng2014,mutualInfoBH,freefall,SchwarzchildHarvestingWellDone}, and being able to describe physically realistic scenarios of local measurements of quantum fields~\cite{Pozas2016,richard,HarvestingDelocalized,carol,jose,boris}. In fact, experimental implementations of the entanglement harvesting protocol are now within reach~\cite{SwitchQEDUpTheLadder,tunableCouplingTowardsHarvesting,cisco2023harvesting}.
    
    In order to model entanglement harvesting, it is common to describe the localized probes as Unruh-DeWitt (UDW) detectors. These are non-relativistic quantum systems which locally couple to quantum fields~\cite{Unruh1976,DeWitt,Schlicht,JormaRigid}. Although UDW models have been shown to accurately describe numerous physical setups~\cite{richard,neutrinos}, they are effective descriptions of bound states of matter that are internally non-relativistic, which can of course be problematic within the context of QFT. Unsurprisingly, the non-relativistic nature of the detectors can cause problems with covariance and causality, so many efforts have been spent in recent years to fully analyze to what extent and in what regimes particle detectors can be used without spoiling the relativistic nature of the QFT they probe~\cite{eduardoOld,us2,PipoFTL,mariaPipoNew}. However, these issues have legitimately raised some questions as to whether analyzing phenomena like entanglement harvesting using internally non-relativistic probes can give trustworthy results. For instance, arguments have been made that perhaps the observed entanglement is an artifact of the non-relativistic nature of the detectors~\cite{max}. 
    
    In order to address these questions, we will use the description presented in~\cite{QFTPD} to replace the non-relativistic particle detector models in the entanglement harvesting protocol by two fully relativistic localized quantum fields that are then used to probe a free quantum field. By considering explicit examples, we show that the two localized fields can extract entanglement from the free field, even if their interaction regions are spacelike separated. This proves that entanglement harvesting is not a consequence of non-localities present because of the non-relativistic nature of commonly employed particle detector models, and that the protocol can be implemented within a fully relativistic framework. We also find that, to leading order in the coupling strength, any two modes of each of the localized QFTs behave exactly like two harmonic oscillator particle detectors, provided suitable choices of coupling regions. This extends the results of~\cite{QFTPD} to setups where multiple detectors are present. Finally we also show that the results obtained with non-relativistic particle detector models provide a lower bound to the results obtained at leading order with fully relativistic QFT probes.

    This manuscript is organized as follows. In Section~\ref{sec:harvesting}, we review the protocol of entanglement harvesting with harmonic oscillator particle detector models. In Section~\ref{sec:QFTPD} we discuss how to use two localized quantum fields in order to locally probe a third quantum field. Section~\ref{sec:fullHarvesting} is devoted to the study of entanglement harvesting using two fully relativistic localized quantum fields coupled to a free field. The conclusions of our work can be found in Section~\ref{sec:conclusions}.

\color{black}

\section{Entanglement Harvesting}\label{sec:harvesting}


It is well known that the vacuum state of a free quantum field contains correlations between different spacetime regions, including those in spacelike separation~\cite{ReehSchlieder,Schlieder1968,vacuumEntanglement,vacuumBell,Haag,WaldQFT,kasiaFewsterIntro}. In fact, in flat spacetimes, for any two spacetime regions, one can always find correlations between local observables of the field. However, when talking about finite regions, it is not so easy to tell whether these correlations come from entanglement or from another form of classical correlations. The reason why addressing this question becomes difficult is the fact that when one effectively restricts a global state of a quantum field theory to finite regions of spacetime, the result is a mixed state, and our techniques for computing entanglement for mixed states are limited, even more so in quantum field theory. One way of approaching the question of how much entanglement there is between two regions of spacetime in a quantum field is to probe the field with simpler localized quantum systems. These systems can couple to the field in the regions of interest in an attempt to extract preexisting entanglement. One can then compute the entanglement between the auxiliary quantum systems in order to infer the entanglement present in the field itself. This is the key idea in the protocol of \textit{entanglement harvesting} (see, e.g.,~\cite{Pozas-Kerstjens:2015}).

The protocol of entanglement harvesting was originally developed in~\cite{Reznik2003,Reznik1}, and has since been refined and studied exhaustively in many different scenarios and spacetimes~\cite{Valentini1991,Reznik2003,Reznik1,reznik2,Salton:2014jaa,Pozas-Kerstjens:2015,Pozas2016,HarvestingSuperposed,Henderson2019,bandlimitedHarv2020,ampEntBH2020,carol,boris,ericksonNew,threeHarvesting2022,twist2022,cisco2023harvesting}. The protocol considers two (or more~\cite{threeHarvesting2022,tripartiteBHarvesting}) particle detectors which couple to the field in distinct regions of spacetime. These particle detectors are usually described as effective non-relativistic models for localized quantum systems that couple to quantum fields (typically called Unruh-DeWitt detectors~\cite{Unruh1976,DeWitt}). Although the main goal of the protocol is for the detectors to extract entanglement that is previously present in the quantum field, harvesting entanglement is not the only way the detectors can get correlated. The field can also establish a communication channel between the two detectors, enabling another possible source of entanglement when the detectors exchange quantum information through the field~\cite{ericksonNew,quantClass}.  In order to isolate the entanglement that is harvested from the field, it is then common to consider detectors whose interaction regions are causally disconnected. This prevents the detectors from exchanging any information\footnote{Notice that the extraction of entanglement by the detectors, including the scenario in which they interact with the field in spacelike separated regions, does not incur into any kind of causality violation. Indeed, the fact that the field vacuum exhibits entanglement between spacelike separated regions is a well-known feature of any relativistic QFT~\cite{vacuumEntanglement,vacuumBell}, and this entanglement, when acquired by the detectors, can never be used to signal between them.}, thus leading to the conclusion that all the entanglement they acquire comes from the field. 

For illustration purposes, let us review a concrete example of entanglement harvesting. We consider two harmonic oscillator particle detectors\footnote{Notice that an analogous derivation follows when a two-level particle detector model is considered instead of a harmonic oscillator. In fact, it can be seen, e.g., in section IV.B of~\cite{Erickson2020entanglementzeromode}, that at leading order a two-level UDW model and a harmonic oscillator model yield exactly the same amount of harvested entanglement when starting in their ground states.} in $3+1$-dimensional Minkowski spacetime, and we label the detectors by $\tc{A}$ and $\tc{B}$. Each detector is assumed to undergo an inertial trajectory $\mf z_\tc{a}(t) = (t, \bm x_\tc{a})$ and $\mf z_{\tc{b}}(t) = (t, \bm x_\tc{b})$. The harmonic oscillator detectors have frequencies $\Omega_\tc{a}$ and $\Omega_\tc{b}$, and their free time evolution is determined by the Hamiltonians
\begin{equation}
    \hat{H}_\tc{a} = \Omega_\tc{a} \hat{a}^\dagger_\tc{a}\hat{a}_{\tc{a}}, \quad\quad 
    \hat{H}_\tc{b} = \Omega_\tc{b} \hat{a}^\dagger_\tc{b}\hat{a}_{\tc{b}},
\end{equation}
where $\hat{a}_\tc{a}$, $\hat{a}^\dagger_\tc{a}$, $\hat{a}_{\tc{b}}$, and $\hat{a}^\dagger_{\tc{b}}$ denote the annihilation and creation operators in the Hilbert space of each detector.

Each harmonic oscillator interacts with the quantum field according to the scalar interaction Hamiltonian densities (or Hamiltonian weights~\cite{us2})
\begin{align}
    \hat{h}_{I,\tc{a}}(\mf x) &= \lambda \left(\Lambda_\tc{a}(\mf x) e^{- \ii \Omega_\tc{a} t} \hat{a}_\tc{a} + \Lambda_\tc{a}^*(\mf x) e^{\ii \Omega_\tc{a} t} \hat{a}^\dagger_\tc{a}\right)\hat{\phi}(\mf{x}),\label{eq:hIA}\\
    \hat{h}_{I,\tc{b}}(\mf x) &= \lambda \left(\Lambda_\tc{b}(\mf x) e^{- \ii \Omega_\tc{b} t} \hat{a}_\tc{b} + \Lambda_\tc{b}^*(\mf x) e^{\ii \Omega_\tc{b} t} \hat{a}^\dagger_\tc{b}\right)\hat{\phi}(\mf{x}),\label{eq:hIB}
\end{align}
where $\Lambda_\tc{a}(\mf x)$ and $\Lambda_{\tc{b}}(\mf x)$ are the spacetime smearing functions, which control the spacetime region where the detectors couple to the field, and $\lambda$ is a coupling constant, which controls the strength of the interactions with the field. The interaction Hamiltonian weight for the full system of the two detectors and field is given by 
\begin{equation}\label{eq:hIABUDW}
    \hat{h}_I(\mf x) = \hat{h}_{I,\tc{a}}(\mf x) + \hat{h}_{I,\tc{b}}(\mf x). 
\end{equation}
The Hamiltonian weight above gives rise to the time evolution operator
\begin{equation}\label{eq:UI}
    \hat{U}_I = \mathcal{T}_t\exp\left( - \ii \int \dd V \hat{h}_I(\mf x)\right),
\end{equation}
where $\mathcal{T}_t\exp$ denotes the time ordering operation with respect to the time parameter $t$ and $\dd V$ is the invariant spacetime volume element. 

Notice that the time ordering operation in Eq. \eqref{eq:UI} privileges the time parameter $t$. This is a consequence of the non-relativistic model employed for the description of the detectors themselves, and is directly linked to the fact that the interaction Hamiltonian weights of Eqs. \eqref{eq:hIA} and \eqref{eq:hIB} violate the microcausality condition whenever the detectors are not pointlike. That is, for spatially smeared detectors, $\hat{h}_{I,\tc{a}}(\mf x)$ and $\hat{h}_{I,\tc{b}}(\mf x)$ do not commute with themselves at spacelike separated points. The incompatibilities of the effective models of particle detectors with relativity are well known~\cite{eduardoOld,us2,PipoFTL,mariaPipoNew}, and are understood to provide a limit for the regime of validity of the models. Nevertheless, these serve as a reminder that particle detector models are an effective description for localized quantum systems, which ultimately should be modelled fully relativistically .

Using the time evolution operator of Eq.~\eqref{eq:UI}, it is then possible to compute the final state of the two detectors after their interaction with the field. For convenience we choose the initial state  \mbox{$\hat{\rho}_0 = \ket{0_\tc{a}}\!\!\bra{0_\tc{a}} \otimes \ket{0_\tc{b}} \!\!\bra{0_{\tc{b}}} \otimes \r_\phi$} for the detectors-field system. That is, we assume that both detectors are initially in their ground states, and that the field is initially in the state $\r_\phi$. We also assume that $\r_\phi$ is a zero-mean Gaussian state. In the regime of weak interactions, it is then possible to use the Dyson expansion for the time evolution operator. The final state of the detectors is found by tracing out the field $\hat{\phi}$. We find that the final state of the detectors to leading order only has components in the subspace spanned by $\{\ket{0_\tc{a}},\ket{1_\tc{a}},\ket{2_\tc{a}}\}\otimes\{\ket{0_\tc{b}},\ket{1_\tc{b}},\ket{2_\tc{b}}\}$. It can be written as
\begin{align}\label{eq:rhoAB}
    \r_\tc{d} = \begin{pmatrix} {\text{\textbf{M}}} & 0_{7\times2} \\ 0_{2\times7} & 0_{2\times2}
\end{pmatrix} +\mathcal{O}(\lambda^4),  
\end{align} 
where
\begin{align}
\text{\textbf{M}} = \begin{pmatrix}
       1- \mathcal{L}_\tc{aa} - \mathcal{L}_\tc{bb}  & 0 & \mathcal{K}_\tc{b}^* & 0 & \mathcal{M}^* & 0 & \mathcal{K}_{\tc{a}}^*\\
        0 & \mathcal{L}_{\textsc{bb}} & 0 & \mathcal{L}_{\textsc{ab}}^* & 0& 0& 0\\
        \mathcal{K}_\tc{b} & 0 & 0 & 0 & 0 & 0& 0\\
        0 & \mathcal{L}_{\textsc{ab}}  & 0 & \mathcal{L}_{\textsc{aa}}  & 0& 0& 0\\
        \mathcal{M} & 0 & 0 & 0 & 0& 0& 0\\
        0 & 0 & 0 & 0 & 0& 0& 0\\
        \mathcal{K}_{\tc{a}} & 0 & 0 & 0 & 0& 0 & 0
    \end{pmatrix},
\end{align}
\begin{align}
    \mathcal{L}_{\tc{ij}} &= \lambda^2 \int \dd V \dd V' \Lambda_\tc{i}(\mf x) \Lambda_\tc{j}(\mf x') e^{- \ii \Omega(t-t')} W(\mf x,\mf x'),\label{eq:Lij}\\
    \mathcal{K}_{\tc{i}} &= \lambda^2 \int \dd V \dd V' \Lambda_\tc{i}(\mf x) \Lambda_\tc{i}(\mf x') e^{\ii \Omega(t+t')} G_F(\mf x,\mf x'),\label{eq:Lij}\\
    \mathcal{M} &= -\lambda^2 \int \dd V \dd V' \Lambda_\tc{a}(\mf x) \Lambda_\tc{b}(\mf x') e^{\ii \Omega(t+t')} G_F(\mf x,\mf x').\label{eq:M}
\end{align}
Here, we have denoted $W(\mf x, \mf x') = \tr_\phi(\hat{\rho}_\phi \hat{\phi}(\mf x)\hat{\phi}(\mf x'))$ and \mbox{$G_F(\mf x, \mf x') = W(\mf x, \mf x')\theta(t-t') + W(\mf x',\mf x) \theta(t'-t)$}, which are the two-point function (or Wightman function) and the Feynman propagator of the field in the state $\hat{\rho}_\phi$. In the expressions above, $\tc{I},\tc{J} \in \{\tc{A},\tc{B}\}$. The $\mathcal{L}_\tc{aa}$, $\mathcal{L}_\tc{bb}$, $\mathcal{K}_\tc{a}$ and $\mathcal{K}_\tc{b}$ terms are local to each detector, while the terms $\mathcal{L}_{\tc{ab}}$, and $\mathcal{M}$ correspond to the correlations acquired by the two detectors.

Our goal is to quantify the entanglement present in the final state of the detectors (if any). Noticing that the final state in Eq. \eqref{eq:rhoAB} is a mixed state (the detectors become entangled with the quantum field), we pick the negativity as an entanglement quantifier. The negativity is an entanglement monotone, which is non-zero only when the state under consideration is entangled. It is defined as the sum of the absolute value of the negative eigenvalues of the partial transpose of a bipartite density operator, and for the state given in Eq.~\eqref{eq:rhoAB}, it reads as
\begin{align}\label{eq:neg}
    \mathcal{N}(\r_\tc{d}) &= \max\left(0,\sqrt{|\mathcal{M}|^2 - \left(\tfrac{\mathcal{L}_\tc{aa} - \mathcal{L}_\tc{bb}}{2}\right)^2} - \tfrac{\mathcal{L}_{\tc{aa}} + \mathcal{L}_\tc{bb}}{2}\right) \nonumber\\
    &\:\:\:\:\:\:\:\:\:\:\:\:\:\:\:\:\:\:\:\:\:\:\:\:\:\:\:\:\:\:\:\:\:\:\:\:\:\:\:\:\:\:\:\:\:\:\:\:\:\:\:\:\:\:\:\:\:\:\:\:\:\:\:\:+ \mathcal{O}(\lambda^4).
\end{align}
Moreover, if the detectors' local terms are the same (i.e., $\mathcal{L}_\tc{aa} = \mathcal{L}_\tc{bb} = \mathcal{L}$), Eq. \eqref{eq:neg} simplifies to \mbox{$\mathcal{N}(\r_\tc{d}) = \max(0,|\mathcal{M}| - \mathcal{L})$}. Overall, the entanglement in the state of the detectors is a competition between the non-local $\mathcal{M}$ term and the local noise terms $\mathcal{L}_\tc{aa}$ and $\mathcal{L}_\tc{bb}$. 

The quantification of how much of the entanglement acquired by the detectors is from communication between them and how much is actual entanglement harvested from the quantum field has been studied in previous literature~\cite{ericksonNew}. If the interaction regions of the detectors are spacelike separated, no entanglement can come from communication. However, many physical examples involve detectors without a finite support (such as atoms~\cite{Pozas2016}), and one has to quantify whether the overlap of the tails of the spacetime smearing functions can generate entanglement originating from communication between the detectors. Means to quantify this are now well known, and it is well understood that if strongly supported spacetime smearing functions are sufficiently separated in space, the communication contribution can be made irrelevant~\cite{ericksonNew,mariaPipoNew}. Using these techniques it is possible to find setups such that (effectively) spacelike separated detectors can harvest entanglement from a quantum field~\cite{Pozas-Kerstjens:2015}. Examples can also be found in the literature where spacelike separated compactly supported spacetime smearing functions can also extract entanglement from a free field~\cite{Pozas-Kerstjens:2015}. 


The entanglement present in spacelike separated regions of spacetime is generally very small, and, while proposals for implementations exist~\cite{tunableCouplingTowardsHarvesting,cisco2023harvesting}, as of today, entanglement harvesting has not yet been experimentally observed. The lack of experimental observations, together with the fact that the protocol has only been studied when considering non-relativistic particle detectors, has raised some questions as to whether entanglement harvesting is an artifact coming from the non-locality of the effective models themselves~\cite{max}. For this reason, one of the goals in this manuscript is to show that it is possible to implement the entanglement harvesting protocol in the context of fully relativistic quantum field theories. Moreover, we will show that the predictions of entanglement harvesting coming from particle detectors are but a lower bound to the amount of leading order entanglement that one can harvest with localized modes of a fully-relativistic quantum field. 

\section{Localized Quantum Fields as Particle Detectors}\label{sec:QFTPD}

In this section we briefly review the setup considered in~\cite{QFTPD}, where localized quantum fields are used to probe a free quantum field theory. We begin the section defining localized fields in Subsection~\ref{sub:localFields}, and then consider the setup in which the localized probe field couples to a Klein Gordon field in Subsection~\ref{sub:QFTPD}.

\subsection{Localized Quantum Fields}\label{sub:localFields}

Consider a scalar field under the influence of a trapping potential $V$ which localizes its modes in space. These examples can be constructed by considering a scalar field $\phi_{\tc d}$ in $3+1$ dimensional Minkowski spacetime with Lagrangian
\begin{equation}\label{eq:lag}
    \mathcal{L} = - \frac{1}{2}\partial_\mu \phi_{\tc d} \partial^\mu \phi_{\tc d} - \frac{m_{\tc{d}}^2}{2} \phi_{\tc d}^2 - V(\bm x) \phi_{\tc d}^2,
\end{equation}
where $m_{\tc{d}}$ is the field's mass, and we employ inertial coordinates $(t,\bm x)$, which we assume to be comoving with the source of the potential, following the approach of~\cite{QFTPD}.

Under the assumption that the potential $V(\bm x)$ is confining (see~\cite{QFTPD} for details), we find that any solution to the equation of motion that originates from the Lagrangian of Eq. \eqref{eq:lag} can be written as a discrete sum of modes:
\begin{equation}\label{eq:fieldmodeexpansion}
    \hat{\phi}_{\tc d}(\mf x) = \sum_{\bm n} \alpha_{\bm n} e^{- \ii \omega_{\bm n}t} \Phi_{\bm n}(\bm x) + \alpha_{\bm n}^* e^{\ii \omega_{\bm n}t} \Phi^*_{\bm n}(\bm x),
\end{equation}
where $\bm n$ is a multi-index, and the functions $\Phi_{\bm n}(\bm x)$ and the eigenfrequencies $\omega_{\bm n}$ are solutions to the eigenvalue problem
\begin{align}
    \left(-\nabla^2 + m^2 + 2V(\bm x)\right)\Phi_{\bm n}(\bm x) = \omega_{\bm n}^2\Phi_{\bm n}(\bm x).
\end{align}

The field can be canonically quantized by promoting the coefficients $\alpha_{\bm n}$ and $\alpha_{\bm n}^*$ to creation and annihilation operators $\hat{a}^\dagger_{\bm n}$ and $\hat{a}_{\bm n}$. This gives rise to the following field representation:
\begin{equation}\label{eq:phidExp}
    \hat{\phi}_{\tc d}(\mf x) = \sum_{\bm n} \hat{a}_{\bm n} e^{- \ii \omega_{\bm n}t} \Phi_{\bm n}(\bm x) + \hat{a}_{\bm n}^\dagger e^{\ii \omega_{\bm n}t} \Phi^*_{\bm n}(\bm x).
\end{equation}
The vacuum state $\ket{0}$ associated with this representation of the field is then the one that satisfies \mbox{$\hat a_{\bm n} \ket{0}=0$} for all $\bm n$. The states of the form $\hat a_{\bm n}^\dagger \ket{0}$ represent (normalized) one-mode excitations of the field with the spatial support of the eigenfunctions $\Phi_{\bm n}(\bm x)$. This means that the presence of the time-independent confining potential allows the quantum field theory we just constructed to have localized states invariant under time translations, which is the main ingredient we need for it to be a sensible model of a realistic probe.


\subsubsection{A field localized by a quadratic potential}\label{sub:harmonicQFT}

We now consider the specific example of a relativistic quantum scalar field in $3+1$ dimensional Minkowski spacetime under the influence of the following quadratic potential:
\begin{equation}
    V(\bm x) = \frac{|\bm x|^2}{2\ell^4}.
\end{equation}
The parameter $\ell$ has dimensions of length and is inversely proportional to the strength of the confining potential. The equation of motion is then given by
\begin{equation}
    \left(\partial_\mu \partial^\mu - m^2 - \frac{|\bm x|^2}{\ell^4}\right) \phi_{\tc d}(\mf x) = 0.
\end{equation}
In order to find a basis of solutions to the equations of motion, we look for solutions of the form $\phi_{\tc d}(\mf x) = e^{-\ii E t} \Phi(\bm x)$, so that the equation turns into an eigenvalue problem:
\begin{equation}\label{eq:EHO}
     \left(E^2 + \nabla^2 - m^2 -\frac{|\bm x|^2}{\ell^4}\right)\Phi(\bm x) = 0.
\end{equation}
The normalizable solutions to this equation are related to the three-dimensional harmonic oscillator, and are found to be
\begin{equation}
    \Phi_{\bm n}(\bm x) = \frac{1}{\sqrt{2\omega_{\bm n}}}f_{n_x}(x)f_{n_y}(y)f_{n_z}(z),
\end{equation}
with $\bm n = (n_x,n_y,n_z)$ being a vector of non-negative integer components, and
\begin{equation}
    f_m(u) = \frac{1}{\sqrt{2^m m!}}\frac{e^{-\frac{u^2}{2\ell^2}}}{\pi^\frac{1}{4} \sqrt{\ell}}H_m(u/\ell),
\end{equation}
where $H_m$ denotes the $m$-th Hermite polynomial. The eigenfrequencies $\omega_{\bm n}$ are given by
\begin{equation}\label{eq:gapHO}
    \omega_{\bm n} = \sqrt{m^2+ \frac{2}{\ell^2}\left(n_x + n_y + n_z +\frac{3}{2}\right)}.
\end{equation}
The corresponding quantum field then admits an expansion as in Eq.~\eqref{eq:phidExp}.

\subsubsection{A field in a cubic cavity}\label{sub:boxQFT}

It is also possible to define a potential which represents a perfectly reflecting cavity. In order to describe a field in a cubic box of side $d$, $U_d = [0,d]^3$, we consider the potential
\begin{equation}\label{eq:potentialCubic}
    V(\bm x) = \begin{cases}
    0, \quad & \bm x \in U_d,\\
    \infty , \quad & \bm x \notin U_d.
    \end{cases}
\end{equation}
This potential makes it so that the field is free inside the box, but is set to zero outside of its walls, effectively implementing Dirichlet boundary conditions. A basis of spatial solutions of the wave equation with this potential can then be written as
\begin{equation}
    \Phi_{\bm n}(\bm x) = \frac{1}{\sqrt{2\omega_{\bm n}} }f_{n_x}(x)f_{n_y}(y)f_{n_z}(z),
\end{equation}
where the functions $f_n(u)$ are given by
\begin{equation}
    f_n(u) = \sqrt{\frac{2}{d}}\sin(\frac{\pi n u}{d}),
\end{equation}
and the corresponding eigenfrequencies are
\begin{equation}\label{eq:gapBox}
    \omega_{\bm n} = \sqrt{m^2 + \frac{\pi^2}{d^2}(n_x^2 + n_y^2 + n_z^2)}.
\end{equation}
The field can be  canonically quantized in the same way as we do for fields under the influence of finite potentials, with the difference that the modes in this case are only non-zero in $U_d$, so that they are compactly supported in space for each $t$.

\subsection{Relativistic Local Probes}\label{sub:QFTPD}

We now focus on the case where the localized field $\hat{\phi}_{\tc{d}}(\mf x)$ is coupled to a free Klein-Gordon field $\hat{\phi}(\mf x)$ which will be the target of the measurement. In this case, the localized field $\hat{\phi}_\tc{d}$ acts as a relativistic probe that can be used to infer properties about the target field $\hat{\phi}$. This setup was detailed in~\cite{QFTPD}, where it was studied in which way each individual mode of the probe field behaves as a smeared harmonic oscillator UDW detector. Here we will briefly review this setup so that we can later apply it to the case where two localized quantum fields couple to a free Klein-Gordon field in Section~\ref{sec:fullHarvesting}.

Let $\phi$ be a free Klein-Gordon field in $3+1$ dimensional Minkowski spacetime, and consider $\phi_\tc{d}$ to be a field localized by a potential $V(\bm x)$, as described in Subsection~\ref{sub:localFields}. We will consider the two fields to be coupled linearly in a finite region of spacetime, so that the system of the two fields is described by the Lagrangian
\begin{equation}\label{eq:lag2}
    \begin{aligned}
    \mathcal{L} =& - \frac{1}{2} \partial_\mu \phi_\tc{d} \partial^\mu \phi_\tc{d} - \frac{m_\tc{d}^2}{2}\phi_{\tc{d}}^2 - V(\bm x) \phi_\tc{d}^2\\
    &- \frac{1}{2} \partial_\mu \phi \partial^\mu \phi - \frac{m^2}{2} \phi^2 - \lambda \zeta(\mf x) \phi_\tc{d} \phi.
    \end{aligned}
\end{equation}
The constants $m_\tc{d}$ and $m$ represent the masses of the fields $\phi_\tc{d}$ and $\phi$, respectively. The function $\zeta(\mf x)$ defines the shape of the spacetime region where the interaction takes place, and $\lambda$ is a coupling constant.

For $\lambda = 0$, the Lagrangian of Eq.~\eqref{eq:lag2} describes two non-interacting fields, each of which can be quantized using standard techniques, with the quantization of the field $\phi_\tc{d}$ being performed as in Subsection~\ref{sub:localFields}. For small values of $\lambda$ one can then treat the two interacting fields perturbatively, using the scalar interaction Hamiltonian density
\begin{equation}
    \hat{h}_I(\mf x) = \lambda \zeta(\mf x) \hat{\phi}_\tc{d}(\mf x) \hat{\phi}(\mf x),
\end{equation}
which generates time evolution in the interaction picture according to the unitary time evolution operator
\begin{equation}\label{eq:UIQFT}
    \hat{U}_I = \mathcal{T}\exp\left(-\ii \int \dd V \hat{h}_I(\mf x)\right).
\end{equation}

We consider the two fields to be completely uncorrelated before the interaction, and we initialize the probe field in its ground state: $\hat{\rho}_0 = \ket{0_\tc{d}}\!\!\bra{0_{\tc{d}}} \otimes \hat{\rho}_\phi$, where $\ket{0_\tc{d}}$ denotes the vacuum state associated to the quantization procedure outlined in Subsection~\ref{sub:localFields}. To make the calculations simpler, we also assume that $\hat\rho_\phi$ is a zero-mean Gaussian state (also called quasifree) of the field $\hat\phi$, although the formalism applies to arbitrary initial states of the target field. This state can then be time evolved using the unitary time evolution operator of Eq.~\eqref{eq:UIQFT}, resulting in the final state
\begin{equation}
    \hat{\rho}_f = \hat{U}_I \hat{\rho}_0 \hat{U}_I^\dagger.
\end{equation}
Since we access the information about the field $\hat\phi$ indirectly by measuring the localized detector field $\hat\phi_\tc{d}$, we trace over the free Klein-Gordon field $\hat\phi$, so that we are left with the state
\begin{equation}
    \hat{\rho}_\tc{d} = \tr_\phi\left(\hat{U}_I \hat{\rho}_0 \hat{U}_I^\dagger \right).
\end{equation}
To leading order in perturbation theory, this state is found to be given by~\cite{QFTPD}
\begin{align}\label{eq:midComputation}
    &\hat{\rho}_\tc{d}=\ket{0_\tc{d}}\!\!\bra{0_\tc{d}}+\lambda^2 \int \dd V \dd V'\zeta(\mf x) \zeta(\mf x')W(\mf x, \mf x') \\
    & \:\:\:\:\:\:\:\:\:\:\:\:\:\:\:\:\:\:\:\:\:\:\:\:\:\:\:\:\:\:\:\:\times \Big(\hat{\phi}_\tc{d}(\mf x')\ket{0_\tc{d}}\!\!\bra{0_\tc{d}}\hat{\phi}_\tc{d}(\mf x) \nonumber\\
    &\:\:\:\:\:\:\:\:\:\:\:\:\:\:\:\:\:\:\:\:\:\:\:\:\:\:\:\:\:\:\:\:\:\:\:\:\:\:\:\:\:-\hat{\phi}_\tc{d}(\mf x)\hat{\phi}_\tc{d}(\mf x')  \ket{0_\tc{d}}\!\!\bra{0_\tc{d}}\theta(t-t') \nonumber\\
    &\:\:\:\:\:\:\:\:\:\:\:\:\:\:\:\:\:\:\:\:\:\:\:\:\:\:\:\:\:\:\:\:\:\:\:\:\:\:\:\:\:\:\:\:\:\:-\ket{0_\tc{d}}\!\!\bra{0_\tc{d}} \hat{\phi}_\tc{d}(\mf x)\hat{\phi}_\tc{d}(\mf x')\theta(t'-t) \Big),\nonumber
\end{align}
where, as before,  $W(\mf x, \mf x') = \tr_\phi(\hat{\phi}(\mf x) \hat{\phi}(\mf x')\hat{\rho}_\phi)$ is the two-point correlation function (or Wightman function) of the quantum field $\hat\phi$ that we aim to probe. 

We can further assume that we only have access to a limited number of modes of the detector field $\hat\phi_\tc{d}$. For simplicity, we will focus on the case where we only have access to one mode of the field, labelled by the multi-index $\bm N$. In this case, it is possible to trace Eq. \eqref{eq:midComputation} over all the other modes of the field, so that we find the final state in the mode $\bm N$ to be~\cite{QFTPD}
\begin{align}
    \r_\N&=\hat{\rho}_{{}_{\!\bm N\!\text{\scriptsize{,0}}}} + \lambda^2 \int \dd V \dd V'W(\mf x, \mf x')
    \Big(\hat{Q}_{{}_{\!\bm N}\!}(\mf x')\hat{\rho}_{{}_{\!\bm N\!\text{\scriptsize{,0}}}}\hat{Q}_{{}_{\!\bm N}\!}(\mf x) \nonumber\\
    &\:\:\:\:\:\:\:\:\:\:\:\:\:\:\:\:\:\:\:\:\:\:\:\:\:\:\:\:\:\:\:\:\:\:\:\:\:\:\:\:\:\:-\hat{Q}_{{}_{\!\bm N}\!}(\mf x)\hat{Q}_{{}_{\!\bm N}\!}(\mf x')  \hat{\rho}_{{}_{\!\bm N\!\text{\scriptsize{,0}}}}\theta(t-t') \nonumber\\
    &\:\:\:\:\:\:\:\:\:\:\:\:\:\:\:\:\:\:\:\:\:\:\:\:\:\:\:\:\:\:\:\:\:\:\:\:\:\:\:\:\:\:\:\:\:\:-\hat{\rho}_{{}_{\!\bm N\!\text{\scriptsize{,0}}}} \hat{Q}_{{}_{\!\bm N}\!}(\mf x)\hat{Q}_{{}_{\!\bm N}\!}(\mf x')\theta(t'-t) \Big)\nonumber\\
    &\:\:\:\:\:\:\:\:\:\:\:\:\:\:\:\:\:\:\:\:\:\:\:\:\:\:\:\:\:\:\:\:\:\:\:\:\:\:\:\:\:\:\:\:\:\:\:\:\:\:\:\:\:\:\:\:\:\:\:\:\:\:\:\:\:\:\:\:\:\:\:+\mathcal{O}(\lambda^4),\label{eq:finalRhoQFTPD}
\end{align}
where $\hat{\rho}_{{}_{\!\bm N\!\text{\scriptsize{,0}}}} = \ket{0_{\N}}\!\!\bra{0_{\N}}$ is the zero occupation number state of the mode $\bm N$, and
\begin{equation}
    \hat{Q}_\N(\mf x)  = \zeta(\mf x) \Phi_\N(\bm x)\left(e^{-\ii \omega_\N t} \hat{a}_\N + e^{\ii \omega_\N t} \hat{a}_\N^\dagger\right)
\end{equation}
is the quadrature operator of the mode $\bm N$ that couples to the quantum field $\hat{\phi}(\mf x)$. Indeed, in~\cite{QFTPD}, it was shown that \textit{to leading order} in the coupling constant, the full result in Eq. \eqref{eq:finalRhoQFTPD} is reproduced by coupling the mode $\bm N$ of $\hat\phi_\tc{d}$ to the target field according to the effective interaction Hamiltonian weight
\begin{equation}
    \hat{h}_\text{eff}(\mf x) = \lambda  \left(\Lambda(\mf x)e^{-\ii \Omega t} \hat{a}_\N + \Lambda^*(\mf x)e^{\ii \Omega t} \hat{a}_\N^\dagger\right) \hat{\phi}(\mf x).
\end{equation}
Here, $\Omega = \omega_\N$ is the frequency of the quantum harmonic oscillator representing mode $\bm N$ of  $\hat\phi_\tc{d}$, and $\Lambda(\mf x)$ is an effective spacetime smearing function, given by the product of the interaction region $\zeta(\mf x)$ and the spatial part of the mode function,
\begin{equation}
    \Lambda(\mf x) = \zeta(\mf x) \Phi_\N(\bm x).
\end{equation}
That is, each individual mode of the field $\hat\phi_\tc{d}$ behaves as a harmonic oscillator particle detector with interaction region determined by the shape of the interaction between the fields and the localization of the mode. 

We remark that the results of this section  can be straightforwardly generalized to include arbitrarily many modes of the probe field as detector degrees of freedom, and for a much more general class of initial states for the probe field not necessarily given by $\ket{0_{\tc{d}}}\bra{0_{\tc{d}}}$. For an explicit verification of this fact using the Schwinger-Keldysh formalism to describe the dynamics of probe and target field, see~\cite{QFTPDPathIntegrals}.

\section{Fully Relativistic Entanglement Harvesting}\label{sec:fullHarvesting}

    In this section we apply the method outlined in Section \ref{sec:QFTPD} to the case where two localized fields are coupled to a free scalar quantum field. In particular, we are interested in the protocol of entanglement harvesting detailed in Subsection~\ref{sec:harvesting}, where two detectors couple to a free quantum field in spacelike separated regions of spacetime in an attempt to extract entanglement present in the field.
    
    Although entanglement harvesting from the vacuum of a quantum field using non-relativistic particle detectors has been studied in a plethora of works in relativistic quantum information, one could reasonably question whether the effective non-relativistic description of the detectors may be contaminating the results. For example, it was argued in~\cite{max} that, as a consequence of the Reeh-Schlieder theorem, localized states of free quantum fields are mixed, so that for any localized detectors, there exists a threshold for the strength of interactions that can entangle localized states. This would not allow local systems to harvest entanglement perturbatively. However, this claim was addressed in~\cite{maxReply}, where it was argued that the argument of~\cite{max} would only apply to specific situations which would generally not correspond to physical setups. In particular, notice that the localized quantum field theories that we considered in Section~\ref{sec:QFTPD} allow for localized pure states. This is a simple example where the arguments of~\cite{max} would not apply. For more details regarding the purity of states of localized QFTs we refer the reader to Appendix~\ref{app:mixed}.
    
    More recently, in~\cite{patricia}, the authors studied entanglement between spatial modes in a quantum field theory, and argued that most individual pairs of spacelike separated modes of a quantum field in $3+1$ dimensions are not entangled. This begs the question of whether non-relativistic particle detectors become entangled as a consequence of their effective description, and not due to entanglement previously present in the quantum field\footnote{In fact, the discussions with the authors of~\cite{patricia} have partially motivated this research. To see how multipartite entanglement of spatial modes of a QFT allow detectors to harvest entanglement we refer to the work in preparation~\cite{patriciaAndUs}.}. 

    Overall, it is not unreasonable to question whether it is possible to extract entanglement from spacelike separated spacetime regions of a quantum field. While the limits of validity of particle detector models are well understood by now, until there is a fully relativistic example of entanglement harvesting, one could still question whether the phenomenon is a mere consequence of the effective theory used for the local probes. To address this, in this section we will show that two localized quantum fields can become entangled by interacting with another free field in spacelike separated interaction regions, thus showcasing a fully relativistic example of entanglement harvesting.

    \subsection{A general framework for entanglement harvesting with localized quantum fields}\label{sub:generalHarvesting}
    
    Consider two localized real scalar quantum fields $\hat{\phi}_\tc{a}(\mf x)$ and $\hat{\phi}_\tc{b}(\mf x)$ in $3+1$ dimensional Minkowski spacetime, under the influence of confining potentials $V_\tc{a}(\bm x)$ and $V_\tc{b}(\bm x)$, according to the description of Subsection~\ref{sub:localFields}. We then consider that there are two vacuum states $\ket{0_\tc{a}}$ and $\ket{0_\tc{b}}$ associated to each field, defined according to the quantization procedure outlined in Section~\ref{sec:QFTPD}. As a consequence of the confining potentials, both fields will have discrete modes labelled by $\bm n_\tc{a}$ and $\bm n_{\tc{b}}$. Each of these fields will linearly interact with a free Klein-Gordon field $\hat{\phi}(\mf x)$, so that the interaction Hamiltonian density of the full theory can be written as
    \begin{equation}
        \hat{h}_I(\mf x) = \lambda  (\zeta_\tc{a}(\mf x)\hat{\phi}(\mf x)\hat{\phi}_\tc{a}(\mf x) + \zeta_\tc{b}(\mf x)\hat\phi(\mf x) \hat{\phi}_\tc{b}(\mf x)),
    \end{equation}
    where $\zeta_\tc{a}(\mf x)$ and $\zeta_{\tc{b}}(\mf x)$ are spacetime smearing functions that are localized in time.

    By picking initial states for the the system of the three fields $\hat{\phi}_\tc{a}(\mf x)$, $\hat{\phi}_\tc{b}(\mf x)$, and $\hat{\phi}(\mf x)$, one can then compute the final state of the system of the probe fields by tracing over $\hat\phi$. In Appendix~\ref{app:twoQFTs} we perform these calculations by considering the initial state \mbox{$\hat{\rho}_0 = \ket{0_\tc{a}}\!\!\bra{0_\tc{a}} \otimes \ket{0_\tc{b}}\!\!\bra{0_\tc{b}} \otimes \hat{\rho}_\phi$}, where $\hat{\rho}_\phi$ is a zero mean Gaussian state for the field $\hat{\phi}(\mf x)$. Moreover, we will assume that we only have access to the modes labelled by $\bm N_\tc{a}$ and $\bm N_\tc{b}$ for the respective fields  $\hat{\phi}_\tc{a}(\mf x)$ and $\hat{\phi}_\tc{b}(\mf x)$. Tracing over all other modes of both fields, we then show (again, in Appendix~\ref{app:twoQFTs}) that, to leading order in perturbation theory, the final state for the modes of the fields is \emph{exactly} the same as one would obtain by considering harmonic oscillator particle detector models. That is, the same final state for the modes can be obtained by considering two quantum harmonic oscillators that interact with the field $\hat{\phi}(\mf x)$ according to the interaction Hamiltonian density
    \begin{equation}\label{eq:hIABQFT}
        \hat{h}_\text{eff}(\mf x) = \lambda \hat{Q}_{{}_{\bm N_\tc{a}}}^{\tc{a}}\!(\mf x) \hat{\phi}(\mf x) +\lambda \hat{Q}_{{}_{\bm N_\tc{b}}}^{\tc{b}}\!(\mf x) \hat{\phi}(\mf x),
    \end{equation}
    where
    \begin{align}
        \hat{Q}_{{}_{\bm N_\tc{a}}}^{\tc{a}}\!(\mf x) = \Lambda_\tc{a}(\mf x) e^{- \ii \Omega_{\tc{a}} t}\hat{a}^{\tc{a}}_{{}_{\bm N_\tc{a}}}+\Lambda^*_\tc{a}(\mf x) e^{\ii \Omega_{\tc{a}} t}\hat{a}^{\tc{a}\dagger}_{{}_{\bm N_\tc{a}}},\\
        \hat{Q}_{{}_{\bm N_\tc{b}}}^\tc{b}\!(\mf x) = \Lambda_\tc{b}(\mf x)e^{- \ii \Omega_{\tc{b}} t} \hat{a}^{\tc{b}}_{{}_{\bm N_\tc{b}}}+\Lambda^\ast_\tc{b}(\mf x) e^{\ii \Omega_{\tc{b}} t}\hat{a}^{\tc{b}\dagger}_{{}_{\bm N_\tc{b}}},
    \end{align}
    with $\hat{a}^{\tc{a}}_{{}_{\bm N_\tc{a}}}$, $\hat{a}^{\tc{a}\dagger}_{{}_{\bm N_\tc{a}}}$, $\hat{a}^{\tc{b}}_{{}_{\bm N_\tc{b}}}$, and $\hat{a}^{\tc{b}\dagger}_{{}_{\bm N_\tc{b}}}$ being the creation and annihilation operators associated with excitations in the modes $\bm N_\tc{a}$ and $\bm N_\tc{b}$ for the respective fields $\hat{\phi}_\tc{a}(\mf x)$ and $\hat{\phi}_\tc{b}(\mf x)$. As in Section \ref{sec:QFTPD}, the spacetime smearing functions are given by the product of $\zeta_{\tc{a}}(\mf x)$ and the spatial mode $\Phi^{\tc{a}}_{{}_{\bm N_\tc{a}}}(\mf x)$ and $\zeta_{\tc{b}}(\mf x)$ and the mode $\Phi^{\tc{b}}_{{}_{\bm N_\tc{b}}}(\mf x)$:
    \begin{align}
        \Lambda_\tc{a}(\mf x) &\coloneqq \zeta_\tc{a}(\mf x) \Phi^{\tc{a}}_{{}_{\bm N_\tc{a}}}(\mf x),\\
        \Lambda_\tc{b}(\mf x) &\coloneqq \zeta_\tc{b}(\mf x) \Phi^{\tc{b}}_{{}_{\bm N_\tc{b}}}(\mf x).
    \end{align}

    In essence, the analogy between particle detectors and modes of localized quantum field theories presented in~\cite{QFTPD} and reviewed in Section \ref{sec:QFTPD} also holds when multiple localized quantum fields are considered. In particular, this means that every result previously studied in entanglement harvesting using particle detectors can be mapped to the case where localized modes of suitable quantum field theories interact with a quantum field. This, in fact, could also be alternatively seen from the path integral approach of~\cite{QFTPDPathIntegrals}, as the arguments used there apply unchanged for any number of modes selected as ``detector'' degrees of freedom, as long as the modes in question are decoupled in the free dynamics of the theory. 

    Interestingly, since at leading order in perturbation theory different modes of the probe fields do not interact with each other, each pair of modes may acquire some amount of entanglement, and not only the two modes labelled by $\bm N_\tc{a}$ and $\bm N_\tc{b}$. This means that the localized fields harvest more entanglement than the amount that a model of two harmonic oscillator particle detectors would predict. Indeed, far from overestimating the amount of entanglement that localized quantum fields can harvest, particle detector models \textit{provide a lower bound} to the amount of leading order entanglement that suitably localized quantum fields would harvest.
    
    An implication of the results above is that it is possible to find fully relativistic quantum field theories such that modes of two localized fields can become entangled after interacting with a free field, even if the interaction regions are spacelike separated. In order to find these fields and modes, it is enough to prescribe potentials and functions $\zeta_\tc{a}(\mf x)$ and $\zeta_\tc{b}(\mf x)$ such that one mode in each field matches the setup considered in any entanglement harvesting protocol considered in the literature. This shows that the protocol of entanglement harvesting can indeed be implemented by fully relativistic theories. 
    
    

    \subsection{Examples}

    For concreteness, we now present specific examples of entanglement harvesting using two localized quantum field theories. Specifically, we consider the lowest energy modes of 1) fields under the influence of quadratic potentials and 2) fields in cubic boxes with Dirichlet boundary conditions, as described in Subsections~\ref{sub:harmonicQFT} and~\ref{sub:boxQFT}. We will see that indeed it is possible for these localized quantum fields to extract entanglement from the vacuum of a free Klein-Gordon field in $3+1$ Minkowski spacetime.

    As our first example of fully relativistic entanglement harvesting, we consider two localized quantum fields in Minkowski spacetime under the influence of potentials $V_\tc{a}(\bm x) = |\bm x|^2/2\ell^4$ and $V_\tc{b}(\bm x) = V_\tc{a}(\bm x - \bm L)$, where $L = |\bm L|$ denotes the proper distance between the centers of the trapping potentials. In essence, the two quantum fields are identical, apart from a spatial shift in the potentials that confine them. Under these assumptions, the energy levels of each field take the form of Eq.~\eqref{eq:gapHO}, with their lowest energy levels being $\omega_{\bm 0_\tc{a}} = \omega_{\bm 0_\tc{b}} = \sqrt{m^2 +   3/\ell^2}$. 
    
    Both fields will interact with a free scalar field $\hat{\phi}(\mf x)$ according to the interaction Hamiltonian of Eq. \eqref{eq:hIABQFT}, where the functions $\zeta_\tc{a}(\mf x)$ and $\zeta_{\tc{b}}(\mf x)$ will conveniently be prescribed as
    \begin{equation}\label{eq:switching}
        \zeta_\tc{a}(\mf x) = \zeta_\tc{b}(\mf x) = e^{-\frac{\pi t^2}{2T^2}}.
    \end{equation}
    This corresponds to interactions that are adiabatically switched on, and peak at $t = 0$. The effective time of the switching is controlled by the timescale $T$. The reason that we consider $\zeta_\tc{a}(\mf x) = \zeta_\tc{b}(\mf x)$ independent of the spatial coordinates is that the effective spacetime region where the localized fields interact with $\hat{\phi}(\mf x)$ is defined by the product of $\zeta_\tc{a}(\mf x)$ with the mode localization of the fields. The spatial localization of the modes together with the time localization of $\zeta_\tc{a}(\mf x)$ then gives an overall interaction which is localized in spacetime for each mode. 

    We consider the three fields to start in their respective ground states, $\ket{0_\tc{a}}\otimes \ket{0_\tc{b}} \otimes \ket{0}$, and we assume that we only have access to the localized fields' mode excitations with the lowest energy, $\omega_{\bm 0_\tc{a}} = \omega_{\bm 0_\tc{b}} \equiv \Omega$. In Appendix~\ref{app:twoQFTs}, we explicitly compute the final state of these modes of the fields, $\hat{\rho}_\tc{d}$, as well as the entanglement in this state as measured by its negativity. In essence, the negativity takes the same form of Eq. \eqref{eq:neg}, and in the case where the excitation probabilities are the same (as we are considering here), it becomes
    \begin{equation}
        \mathcal{N}(\r_\tc{d}) = \max(0, |\mathcal{M}| - \mathcal{L}) + \mathcal{O}(\lambda^4),
    \end{equation}
    where the $\mathcal{L}$ and $\mathcal{M}$ terms are given by
    \begin{align}
        \mathcal{L} &= \lambda^2 \int \dd V \dd V' \Lambda_\tc{a}(\mf x)\Lambda_\tc{a}(\mf x') e^{- \ii \Omega(t-t')} W(\mf x,\mf x'),\label{eq:LQFTA}\\
        &= \lambda^2 \int \dd V \dd V' \Lambda_\tc{b}(\mf x)\Lambda_\tc{b}(\mf x') e^{- \ii\Omega (t-t')} W(\mf x,\mf x'),\nonumber\\
        \mathcal{M} &= -\lambda^2 \int \dd V \dd V' \Lambda_\tc{a}(\mf x)\Lambda_\tc{b}(\mf x') e^{\ii \Omega(t+t')} G_F(\mf x,\mf x'),\nonumber
    \end{align}
    with $\Omega = \sqrt{m^2 + 3/\ell^2}$ and the spacetime smearing functions are given by
    \begin{align}
        \Lambda_\tc{a}(\mf x) &= \zeta_\tc{a}(\mf x) \Phi^{\tc{a}}_{\bm 0_\tc{a}}\!(\bm x) = e^{-\frac{\pi t^2}{2T^2}} \!\!\left(\frac{1}{\pi\ell^2}\right)^{\frac{3}{4}}\!\!\!\frac{e^{-\frac{|\bm x|^2}{2\ell^2}}}{\left(m^2 + \tfrac{3}{\ell^2}\right)^{1/4}},\nonumber\\
        \Lambda_\tc{b}(\mf x) &= \zeta_\tc{b}(\mf x) \Phi^{\tc{b}}_{\bm 0_\tc{b}}\!(\bm x) = \Lambda_\tc{a}(t,\bm x - \bm L).\label{eq:smearingsQFTHarvesting}
    \end{align}
    We then see that the effective size of the interaction region can be estimated by looking at the standard deviation of the space dependent Gaussian function in Eq.~\eqref{eq:smearingsQFTHarvesting}. In this case, the spatial size of the interaction region can be estimated to be $\sigma \sim \ell$, so that smaller values of the parameter $\ell$ that defines the confining potential corresponds to more localized detectors.

    \begin{figure}[h!]
        \centering
        \includegraphics[width=8.6cm]{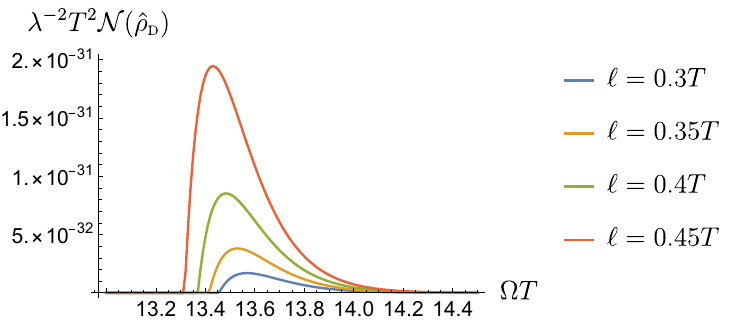}
        \caption{The negativity of the state of two localized quantum fields confined by a quadratic potential when restricted to their lowest energy after interacting with a massless scalar field. The negativity is plotted as a function of the energy of the modes $\Omega = \sqrt{m^2 + 3/\ell^2}$. The time duration of the interaction $T$ is used as a scale. The separation between the detectors interaction regions for these plots is $L = 5 T$.}
        \label{fig:quadratic}
    \end{figure}
    
   Motivated by the discussion of entanglement harvesting, we focus on the case where the interaction regions are approximately spacelike separated\footnote{Even though the Gaussian tails of the switching are technically infinitely long, using these switchings is effectively equivalent to considering compactly supported switchings. For more details about effective communication from Gaussian tails and how the use of Gaussians is justified, we refer the reader to~\cite{ericksonNew,mariaPipoNew}.}, so that entanglement acquired by the localized modes via communication can be neglected. For this reason we consider $L = 5T$ in the specific example that we explore here, where we verified that the effect of communication is $6$ orders of magnitude smaller than the effects arising from vacuum entanglement harvesting~\cite{ericksonNew,quantClass}. In Fig.~\ref{fig:quadratic} we plot the entanglement acquired by the localized modes as a function of their energy gap. The plot is what is expected for the behaviour of entanglement harvesting in the Minkowski vacuum, where there is a threshold in the energy gap below which no entanglement can be extracted. For $\Omega T$ above this threshold, the entanglement peaks and quickly decays. 
    The behaviour seen in Fig.~\ref{fig:quadratic} is identical to that of a UDW detector with a Gaussian spacetime smearing function, as expected (for comparison, see e.g.~\cite{hectorMass}).

    \begin{figure}[h!]
        \centering
        \includegraphics[width=8.6cm]{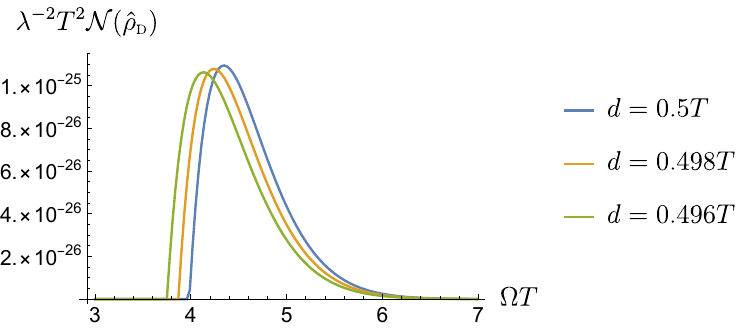}
        \caption{The negativity of the state of two localized quantum fields in boxes of sides $d$ when restricted to their lowest energy mode after spacelike interaction with a massless scalar field. The negativity is plotted as a function of the energy of the localized mode, $\Omega = \sqrt{m^2 + 3\pi/d^2}$. The time duration of the interaction $T$ is used as a scale. The separation between the detectors interaction regions for these plots is $L = 4.5 T$.}
        \label{fig:box}
    \end{figure}

    Finally, in Fig.~\ref{fig:box} we consider the case where two massive fields in cubic cavities of size length $d$ with Dirichlet boundary conditions interact with a free massless scalar field. The box localization was discussed in Subsection~\ref{sub:boxQFT}. We consider the same choices of $\zeta_\tc{a}(\mf x)$ and $\zeta_\tc{b}(\mf x)$  as in Eq. \eqref{eq:switching}. We also restrict the localized fields to the lowest energy mode $\bm 1_\tc{a} = \bm 1_{\tc b} = (1,1,1)$, with energy $\omega^\tc{a}_{\bm 1_\tc{a}} = \omega^{\tc{b}}_{\bm 1_{\tc{b}}} = \sqrt{m^2 + 3\pi^2/d^2}$. To ensure that communication between the detectors is negligible, we pick $d \sim 0.5 T$ and consider the distance between the cavities to be given by $L = 5T$. The negativity in this case can be seen in Fig.~\ref{fig:box} as a function of the energy gap of the $\bm 1_{\tc{a}}$, $\bm 1_{\tc{b}}$ modes of the fields. The behaviour of the negativity is similar to most cases of entanglement harvesting in spacelike separated regions. We see more entanglement in this setup due to the smaller choice of $L$, which can be taken in this case because the communication between the detectors is naturally smaller due to the compact support of the modes in space.

    Figs.~\ref{fig:quadratic} and~\ref{fig:box} showcase examples where two completely relativistic (microcausal) localized quantum field theories can extract entanglement from the vacuum of a free field. This proves that entanglement harvesting is not a consequence of any features of effective non-relativistic theories, and it is indeed a prediction of quantum field theory.



\subsection{Entanglement harvesting with one probe field}

It is also possible to model entanglement harvesting with two effectively localized particle detectors that both emerge from a single probe quantum field. This is conceptually closer to situations where the probes  whose entanglement we wish to test at the end of the experiment are identical---such as,  for instance, using two electrons held in two distinct positions as probes of the electromagnetic field. In fact, it is relatively straightforward to see how this can be achieved by noting that Section~\ref{sec:QFTPD} immediately generalizes to the case where the confining potential may actually localize the field in multiple regions instead of just one.

It is indeed possible to treat a field defined in a set with multiple connected components $U = U_1 \cup ... \cup U_n$ using the formalism of Section~\ref{sec:QFTPD} in each connected component $U_i$. Indeed, when multiple connected components are present, the equations of motion in each of them decouple, effectively giving rise to different quantum fields $\hat{\phi}_i(\mf x)$ defined in each of the sets $U_i$, as the Hilbert space for the theory factors are a tensor product of non-interacting fields localized in each region $U_i$. 

This can be done in an approximate sense even when one does not strictly have multiple connected components but instead the potential yields strongly supported modes in two sufficiently separated regions---for example, if the potential has two regions of minima separated by a sufficiently large potential wall. In this case the probe field approximately factors as two non-interacting infinite towers of harmonic oscillators localized around two distinct regions of space, each of which could equivalently be treated as emerging from a quantum field of its own.

For concreteness, let us assume that the local minima of the potential $V(\bm{x})$ are found in two finite spatial regions $R_{\tc{a}}$ and $R_{\tc{b}}$, and that the potential goes to infinity as we move away from both regions. Say that, in $R_{\tc{a}}$ and $R_{\tc{b}}$, the potential locally behaves as $V_{\tc{a}}(\bm{x})$ and $V_{\tc{b}}(\bm{x})$ respectively, where $V_{\tc{a}}(\bm{x})$ and $V_{\tc{b}}(\bm{x})$ are both confining potentials of their own. Now, the field can be approximately split as
\begin{equation}
    \hat{\phi}_{\tc{d}}(\mf x) \simeq \hat{\phi}_\tc{a}(\mf x) + \hat{\phi}_\tc{b}(\mf x),
\end{equation}
where both $\hat{\phi}_\tc{a}(\mf x)$ and $\hat{\phi}_\tc{b}(\mf x)$ can be written in a mode expansion as in Eq. \eqref{eq:phidExp}, with the spatial mode functions $\Phi^{\tc{a,b}}_{\bm n_\tc{a,b}}(\bm{x})$ corresponding to spatially localized profiles associated to the potentials $V_{\tc{a,b}}(\bm{x})$ and localized around regions $R_{\tc{a,b}}$ respectively. The assumption that the potential barrier and spatial separation between the regions $R_{\tc{a}}$ and $R_{\tc{b}}$ are sufficiently large then translates to the condition that the overlap between any two modes associated to distinct regions is negligible, i.e., 
\begin{equation}\label{eq:nooverlapassumption}
    \int \dd^d \bm{x}\,\Phi^{\tc{a}}_{\bm n_{\tc{a}}}(\bm{x})\Phi^{\tc{b}}_{\bm{n}_{\tc{b}}}(\bm{x}) \simeq 0 \,\,\,\,\forall \,\,\bm n_{\tc{a}}, \bm n_{\tc{b}}.
\end{equation}

We can interpret this as showing that, when the potential yields strongly supported modes in two sufficiently separated regions, the probe field essentially factors as two non-interacting infinite towers of harmonic oscillators localized around two distinct regions of space, each of which could equivalently be treated as emerging from a distinct quantum field. The condition given by Eq.~\eqref{eq:nooverlapassumption} guarantees that the creation and annihilation operators of the modes supported in each separate region approximately commute. The vacuum state for the probe field can then be approximated as a tensor product of the local vacua of the independent theories on each localization region. In summary, within these approximations, this procedure is mathematically the same as quantizing two independent fields, each under the influence of a different trapping potential with only one minimum in either $R_{\tc{a}}$ or $R_{\tc{b}}$.  This shows how one can conceptualize entanglement harvesting setups where each particle detector is obtained as a different localized mode of one single quantum field. From this point onwards, the math that led to the results in Figs.~\ref{fig:quadratic} and~\ref{fig:box} remains unchanged.

\section{Conclusions}\label{sec:conclusions}


    We studied the protocol of entanglement harvesting in the case where the probes harvesting entanglement from the field are modelled as localized fully relativistic quantum fields themselves, in contrast to the commonly employed treatment where the probes are modelled as non-relativistic particle detectors. 
    
    In particular, we found examples where modes of the two localized probe fields become entangled after interacting with a free Klein-Gordon field when the regions of interaction of each field are spacelike separated. This showcases a protocol for extracting spacelike entanglement from the vacuum of a free quantum field where not only the fields, but also the probes are fully relativistic, contrary to the intuitions derived in~\cite{max}.  

    We also showed how to reduce two localized QFT probes to particle detector models, extending the results of~\cite{QFTPD} to situations where more than one localized field is probing a free QFT. Indeed, we verified that the commonly employed particle detector models provide an accurate description for these scenarios, approximating the fully relativistic results at leading order.
    
    Moreover, we showed that---at leading order in perturbation theory---the entanglement that the fully relativistic probes can harvest is actually bounded from below by the results obtained when the probes are harmonic-oscillator Unruh-DeWitt detectors, further legitimizing the use of particle detector models to study this kind of protocols.





\begin{acknowledgements}
    The authors thank Christopher J. Fewster for valuable discussions. TRP acknowledges support from the Natural Sciences and Engineering Research Council of Canada (NSERC) through the Vanier Canada Graduate Scholarship. JPG acknowledges the support of a fellowship from “La Caixa” Foundation (ID 100010434, with fellowship code LCF/BQ/AA20/11820043). JPG and BSLT acknowledge support from the Mike and Ophelia Lazaridis Fellowship.  EMM acknowledges support through the Discovery Grant Program of the Natural Sciences and Engineering Research Council of Canada (NSERC). EMM also acknowledges support of his Ontario Early Researcher award. Research at Perimeter Institute is supported in part by the Government of Canada through the Department of Innovation, Science and Industry Canada and by the Province of Ontario through the Ministry of Colleges and Universities. 
\end{acknowledgements}

\onecolumngrid
\appendix

\section{Two Fields in Cavities interacting with a K.G. field}\label{app:twoQFTs}

Denote the field in each cavity by $\hat{\phi}_\tc{a}$ and $\hat{\phi}_\tc{b}$, and the free field that they both interact with by $\hat{\phi}$ in the regions defined by the supports of $\zeta_\tc{a}(\mf x)$ and $\zeta_\tc{b}(\mf x)$, respectively. The interaction Hamiltonian density of the theory will be prescribed as
\begin{equation}
    \hat{h}_I(\mf x) = \lambda  (\zeta_\tc{a}(\mf x)\hat{\phi}(\mf x)\hat{\phi}_\tc{a}(\mf x) + \zeta_\tc{b}(\mf x)\phi(\mf x) \hat{\phi}_\tc{b}(\mf x)).
\end{equation}
We now write the $\phi_{\tc{a},\tc{b}}$ fields with the spacetime smearing functions as
\begin{align}
\zeta_\tc{a}(\mf x)\hat{\phi}_\tc{a}(\mf x) &= \sum_{\bm n} \zeta_\tc{a}(\mf x)\left( u_{\bm n}^{\tc{a}}(\mf x) \hat{a}_{\bm n}^\tc{a} +  u^{\tc{a}*}_{\bm n}(\mf x)\hat{a}^{\tc{a}\dagger}_{\bm n} \right) = \sum_{\bm n} \hat{Q}^{\tc{a}}_{\bm n}(\mf x)\\
\zeta_\tc{b}(\mf x)\hat{\phi}_\tc{b}(\mf x) &= \sum_{\bm n} \zeta_\tc{b}(\mf x)\left( u_{\bm n}^{\tc{b}}(\mf x) \hat{a}_{\bm n}^\tc{b} +  u^{\tc{b}*}_{\bm n}(\mf x) \hat{a}_{\bm n}^{\tc{b}\dagger} \right) = \sum_{\bm n}   \hat{Q}_{\bm n}^\tc{b}(\mf x),
\end{align}
where $u_{\bm n}^{\tc{a}}(\mf x) = e^{-\ii \omega_{\bm n}^{\tc{a}} t}\Phi^\tc{a}_{\bm n}(\bm x)$, $u_{\bm n}^{\tc{b}}(\mf x) = e^{-\ii \omega_{\bm n}^{\tc{b}} t}\Phi_{\bm n}^{\tc{b}}(\bm x)$, and the field expansion will depend on the specific boundary conditions and equations of motion. We are working under the assumption that the field has discrete energy levels, so that the sums above are discrete. The field expansions automatically define states $\ket{0_\tc{a}}$ and $\ket{0_\tc{b}}$, which are annihilated by all operators $\hat{a}_{\bm n}^\tc{a}$ and $\hat{a}_{\bm n}^\tc{b}$, respectively. We can then write the Hamiltonian interaction density as
\begin{equation}
    \hat{h}_I(\mf x) = \lambda\hat{\phi}(\mf x) \sum_{\bm n} \left(\hat{Q}_{\bm n}^{\tc{a}}(\mf x) +  \hat{Q}_{\bm n}^\tc{b}(\mf x)\right).
\end{equation}
In perturbation theory, we then get
\begin{equation}
    \hat{U} = \mathcal{T}\exp(- \ii \int \dd V \hat{h}_I(\mf x)) = \openone + \hat{U}^{(1)} + \hat{U}^{(2)} + \mathcal{O}(\lambda^3),
\end{equation}
where:
\begin{align}
    \hat{U}^{(1)} = - \ii \int \dd V \hat{h}_I(\mf x) = - \ii \lambda \int \dd V  \hat{\phi}(\mf x)  \sum_{\bm n} \left(\hat{Q}^\tc{a}_{\bm n}(\mf x) +  Q_{\bm n}^\tc{b}(\mf x)\right).
\end{align}
\begin{align}
    \hat{U}^{(2)}& =  - \int \dd V\dd V' \hat{h}_I(\mf x) \hat{h}_I(\mf x')\theta(t-t')\\&= - \int \dd V \dd V' \hat{\phi}(\mf x)\hat{\phi}(\mf x')\theta(t-t') \sum_{nm} \Big( \hat{Q}_{\bm n}^{\tc{a}}(\mf x)\hat{Q}_{\bm m}^{\tc{a}}(\mf x')+ \hat{Q}^\tc{b}_{\bm n}(\mf x)\hat{Q}_{\bm m}^{\tc{b}}(\mf x')+  \hat{Q}^\tc{a}_{\bm n}(\mf x)\hat{Q}^\tc{b}_{\bm m}(\mf x') +  \hat{Q}^\tc{b}_{\bm n}(\mf x) \hat{Q}^\tc{a}_{\bm m}(\mf x')\Big).
\end{align}
The final state will be given by
\begin{equation}
    \hat{\rho}_f = \hat{U}\hat{\rho}_0\hat{U}^\dagger = \hat{\rho}_0 + \hat{U}^{(1)}\hat{\rho}_0 + \hat{\rho}_0 \hat{U}^{(1)\dagger} + \hat{U}^{(1)}\hat{\rho}_0\hat{U}^{(1)\dagger}+ \hat{U}^{(2)}\hat{\rho}_0 + \hat{\rho}_0 \hat{U}^{(2)\dagger} + \mathcal{O}(\lambda^3).
\end{equation}
We will assume that $\hat{\rho}_0 = \ket{0_\tc{a}0_\tc{b}}\!\!\bra{0_\tc{a}0_\tc{b}}\otimes \hat{\rho}_\phi = \hat{\rho}_{0,\tc{ab}}\otimes \hat{\rho}_\phi$ and that $\tr_\phi(\hat{U}^{(1)}\hat{\rho}_0) = 0$, so that the $\mathcal{O}(\lambda)$ terms do not contribute to the partial state of the cavities A and B. We then only need to compute \mbox{$\tr_\phi(\hat{U}^{(1)}\hat{\rho}_0\hat{U}^{(1)\dagger}+ \hat{U}^{(2)}\hat{\rho}_0 + \hat{\rho}_0 \hat{U}^{(2)\dagger})$}. We have:

\begin{align}
    &\tr_\phi(\hat{U}^{(1)}\hat{\rho}_0\hat{U}^{(1)\dagger}) \nonumber\\&= \lambda^2\int \dd V\dd V' W(\mf x', \mf x)  \sum_{nm} \Big( \hat{Q}^{\tc{a}}_{\bm n}(\mf x)\hat{\rho}_{0,\tc{ab}}\hat{Q}^{\tc{a}}_{\bm m}(\mf x')+ \hat{Q}^\tc{b}_{\bm n}(\mf x)\hat{\rho}_{0,\tc{ab}}\hat{Q}_{\bm m}^{\tc{b}}(\mf x')+ \hat{Q}^\tc{a}_{\bm n}(\mf x)\hat{\rho}_{0,\tc{ab}}\hat{Q}^{\tc{b}}_{\bm m}(\mf x') +  \hat{Q}^\tc{b}_{\bm n}(\mf x) \hat{\rho}_{0,\tc{ab}}\hat{Q}^{\tc{a}}_{\bm m}(\mf x')\Big),
\end{align}
and
\begin{align}
    \tr_\phi(\hat{U}^{(2)}\hat{\rho}_0) &= - \lambda^2\int \dd V \dd V' W(\mf x, \mf x')\theta(t-t')  \nonumber\\\times&\sum_{nm} \Big( \hat{Q}^{\tc{a}}_{\bm n}(\mf x)\hat{Q}^{\tc{a}}_{\bm m}(\mf x')\hat{\rho}_{0,\tc{ab}}+ \hat{Q}^\tc{b}_{\bm n}(\mf x)\hat{Q}_{\bm m}^{\tc{b}}(\mf x')\hat{\rho}_{0,\tc{ab}}+  \hat{Q}^\tc{a}_{\bm n}(\mf x)\hat{Q}^\tc{b}_{\bm m}(\mf x')\hat{\rho}_{0,\tc{ab}} +  \hat{Q}^\tc{b}_{\bm n}(\mf x) \hat{Q}^\tc{a}_{\bm m}(\mf x')\hat{\rho}_{0,\tc{ab}}\Big),
\end{align}
where $W(\mf x, \mf x') = \text{tr}(\hat{\rho}_\phi \hat{\phi}(\mf x) \hat{\phi}(\mf x'))$. Notice that 
\begin{equation}
    \hat{\rho}_{0,\tc{ab}} = \ket{0_\tc{a}0_\tc{b}}\!\!\bra{0_\tc{a}0_\tc{b}} = \bigotimes_{nm} \ket{0_{\bm n}^\tc{a}0_{\bm m}^\tc{b}}\!\!\bra{0_{\bm n}^\tc{a}0_{\bm m}^\tc{b}} = \hat{\rho}_{\tc{d},0}\bigotimes_{n,m>1} \ket{0_{\bm n}^\tc{a}0_{\bm m}^\tc{b}}\!\!\bra{0_{\bm n}^\tc{a}0_{\bm m}^\tc{b}},
\end{equation}
where $\ket{0_{\bm n}^{\tc{a}}}$ and $\ket{0_{\bm n}^{\tc{b}}}$ denotes the ground states of each harmonic of each cavity, and 
\begin{equation}
    \hat{\rho}_{\tc{d},0} = \ket{0_1^\tc{a}0_1^\tc{b}}\!\!\bra{0_1^\tc{a}0_1^\tc{b}}
\end{equation}
is the ground state of the first harmonic in each cavity. We then trace out all cavity modes except for the first harmonic. That is, we will compute the density operator \mbox{$\tr_{\phi,H}(\hat{\rho}_f) = \tr_H(\tr_\phi(\hat{U}^{(1)}\hat{\rho}_0\hat{U}^{(1)\dagger}+ \hat{U}^{(2)}\hat{\rho}_0 + \hat{\rho}_0 \hat{U}^{(2)\dagger}))$}, where $\tr_H$ denotes the trace over all cavity harmonics, except the first, for fields A and B.

Overall, we will need to compute the trace of quantities of the form $\tr_H(\hat{Q}^\tc{a}_{\bm n}(\mf x)\hat{Q}^\tc{a}_{\bm m}(\mf x')\hat{\rho}_{0,\tc{ab}})$. We know that since $\hat{Q}^\tc{i}_{\bm n}(\mf x) = (u_{\bm n}^{\tc{i}}(\mf x)\hat{a}_{\bm n}^\tc{i} + u_{\bm n}^{\tc{i}*}(\mf x)\hat{a}^{\tc{i}\dagger}_{\bm n})$ for $\tc{I} = \tc{A},\tc{B}$. Therefore, the products of the form $\hat{Q}^\tc{i}_{\bm n}(\mf x)\hat{Q}_{\bm m}^\tc{i}(\mf x')$ will only give non-diagonal elements if $n = m$. When $n= m \neq 1$, we have:
\begin{equation}
    \tr_\tc{a}(\hat{Q}^\tc{a}_{\bm n}(\mf x)\hat{Q}^\tc{a}_{\bm n}(\mf x')\ket{0^\tc{a}_{\bm n}}\!\!\bra{0^\tc{a}_{\bm n}}) = \bra{0_{\bm n}^\tc{a}}\hat{Q}^\tc{a}_{\bm n}(\mf x)\hat{Q}^\tc{a}_{\bm n}(\mf x')\ket{0_{\bm n}^\tc{a}} = \zeta_\tc{a}(\mf x)\zeta_\tc{a}(\mf x')u_{\bm n}^\tc{a}(\mf x) u_{\bm n}^{\tc{a}*}(\mf x')
\end{equation}
and
\begin{equation}
    \tr_\tc{b}(\hat{Q}^\tc{b}_{\bm n}(\mf x)\hat{Q}^\tc{b}_{\bm n}(\mf x')\ket{0^\tc{b}_{\bm n}}\!\!\bra{0^\tc{b}_{\bm n}}) = \bra{0_{\bm n}^\tc{b}}\hat{Q}^\tc{b}_{\bm n}(\mf x)\hat{Q}^\tc{b}_{\bm n}(\mf x')\ket{0_{\bm n}^\tc{b}} = \zeta_\tc{b}(\mf x) \zeta_\tc{b}(\mf x')u_{\bm n}^\tc{b}(\mf x) u_{\bm n}^{\tc{b}*}(\mf x').
\end{equation}
We find that for $n$ and $m$ different from $1$,
\begin{equation}
    \tr_H(\hat{Q}_{\bm n}^\tc{i}(\mf x)\hat{Q}_{\bm m}^\tc{j}(\mf x')\hat{\rho}_{0,\tc{ab}}) = \delta_{\bm{nm}} \delta_{\tc{i}\tc{j}} \zeta_\tc{i}(\mf x)\zeta_\tc{j}(\mf x')u_{\bm n}^{\tc{i}}(\mf x)u_{\bm m}^{\tc{j}*}(\mf x')\hat{\rho}_{\tc{d},0},
\end{equation}
where $\tr_H(\hat{Q}_{\bm n}^\tc{a}(\mf x)\hat{Q}_{\bm m}^\tc{b}(\mf x')\hat{\rho}_{0,\tc{ab}}) = 0$ automatically, as it factors into expectation values of creation and annihilation operators in $\tc{A}$ and $\tc{B}$ evaluated at the vacuum. Finally, notice that when $n = m = 1$ we do not need to trace over it, because $H$ encompasses every harmonic except for the first one.

Putting the results above together, we then find that
\begin{align}
    \tr_H(\tr_\phi(\hat{U}^{(1)} \hat{\rho}_0\hat{U}^{(1)\dagger})) = \!\lambda^2\!\!\!\int \!\dd V \dd V'  W(\mf x' ,\mf x)\Big(& \hat{Q}^{\tc{a}}_1(\mf x)\hat{\rho}_{\tc{d},0}\hat{Q}_1^{\tc{a}}(\mf x')+ \hat{Q}_1^\tc{b}(\mf x)\hat{\rho}_{\tc{d},0}\hat{Q}_1^{\tc{b}}(\mf x')+ \hat{Q}^\tc{a}_1(\mf x)\hat{\rho}_{\tc{d},0}\hat{Q}^{\tc{b}}_1(\mf x') +  \hat{Q}_1^\tc{b}(\mf x) \hat{\rho}_{\tc{d},0}\hat{Q}^{\tc{a}}_1(\mf x') \nonumber\\&+\sum_{n>1}( \zeta_\tc{a}(\mf x)\zeta_\tc{a}(\mf x')u_{\bm n}^{\tc{a}}(\mf x) u_{\bm n}^{\tc{a}*}(\mf x')+\zeta_\tc{b}(\mf x)\zeta_\tc{b}(\mf x')u_{\bm n}^{\tc{b}}(\mf x) u_{\bm n}^{\tc{b}*}(\mf x'))\hat{\rho}_{\tc{d},0}\Big)
\end{align}
and
\begin{align}
    \!\tr_H(\tr_\phi(\hat{U}^{(2)}\hat{\rho}_0))\! =\! - \lambda^2\!\!\!\int\! \!\dd V \dd V' W(\mf x, \mf x')\theta(t-t')\! \Big( &\hat{Q}_1^{\tc{a}}(\mf x)\hat{Q}_1^{\tc{a}}(\mf x')\hat{\rho}_{\tc{d},0}\!+\! \hat{Q}^\tc{b}_1(\mf x)\hat{Q}_1^{\tc{b}}(\mf x')\hat{\rho}_{\tc{d},0}\!+\! \hat{Q}^\tc{a}_1(\mf x)\hat{Q}^\tc{b}_1(\mf x')\hat{\rho}_{\tc{d},0}\! +\! \hat{Q}^\tc{b}_1(\mf x) \hat{Q}^\tc{a}_1(\mf x')\hat{\rho}_{\tc{d},0}\nonumber\\
   & +\sum_{n>1} (\zeta_\tc{a}(\mf x)\zeta_\tc{a}(\mf x')u_{\bm n}^{\tc{a}}(\mf x)  u_{\bm n}^{\tc{a}*}(\mf x')+ \zeta_\tc{b}(\mf x)\zeta_\tc{b}(\mf x')u_{\bm n}^{\tc{b}}(\mf x) u_{\bm n}^{\tc{b}*}(\mf x'))\hat{\rho}_{\tc{d},0}\Big).
   \end{align}
The last term $\tr_H(\tr_\phi(\hat{\rho}_0 \hat{U}^{(2)\dagger}))$ is simply the conjugate of the term above. Notice that the terms proportional to $\hat{\rho}_{\tc{d},0}$ will cancel when all terms are added together. This can be seen from explicit calculation using $\theta(t-t') + \theta(t'-t) = 1$, or simply by noticing that each term in the Dyson expansion is traceless due to trace preservation. 

The products of terms $\hat{Q}_1^\tc{i}(\mf x)\hat{Q}_1^\tc{j}(\mf x')$ is given by

\begin{align}
    \hat{Q}_1^\tc{i}(\mf x)\hat{Q}_1^\tc{j}(\mf x') &= \zeta_{\tc{i}}(\mf x)\zeta_{\tc{j}}(\mf x') (u_1^\tc{i}(\mf x) \hat{a}_{\bm n}^\tc{i} + u_1^{\tc{i}*}(\mf x) \hat{a}_{\bm n}^{\tc{i}\dagger}) (u_1^\tc{j}(\mf x') \hat{a}_{\bm n}^\tc{j} + u_1^{\tc{j}*}(\mf x') \hat{a}_{\bm n}^{\tc{j}\dagger}) \\
    &= \zeta_{\tc{i}}(\mf x)\zeta_{\tc{j}}(\mf x') (u_1^\tc{i}(\mf x) u_1^\tc{j}(\mf x') \hat{a}_{\bm n}^\tc{i}\hat{a}_{\bm n}^\tc{j} + u_1^{\tc{i}*}(\mf x)u_1^\tc{j}(\mf x') \hat{a}_{\bm n}^{\tc{i}\dagger}\hat{a}_{\bm n}^\tc{j} + u_1^\tc{i}(\mf x)u_1^{\tc{j}*}(\mf x') \hat{a}_{\bm n}^\tc{i} \hat{a}_{\bm n}^{\tc{j}\dagger} + u_1^{\tc{i}*}(\mf x) u_1^{\tc{j}*}(\mf x') \hat{a}_{\bm n}^{\tc{i}\dagger}\hat{a}_{\bm n}^{\tc{j}\dagger}).
\end{align}
We then define the spacetime smearing functions
\begin{align}
    \Lambda_\tc{a}(\mf x) &\coloneqq \zeta_\tc{a}(\mf x) u^{\tc{a}}_1(\mf x),\\
    \Lambda_\tc{b}(\mf x) &\coloneqq \zeta_\tc{b}(\mf x) u^{\tc{b}}_1(\mf x),    
\end{align}
so that the $\hat{Q}_1^{\tc{i}}(\mf x)$ terms read simply as
\begin{align}
    \hat{Q}_1^{\tc{a}}(\mf x) = \Lambda_\tc{a}(\mf x) \hat{a}^{\tc{a}}_1+\Lambda^*_\tc{a}(\mf x) \hat{a}^{\tc{a}\dagger}_1,\\
    \hat{Q}_1^{\tc{b}}(\mf x) = \Lambda_\tc{b}(\mf x) \hat{a}^{\tc{b}}_1+\Lambda^*_\tc{b}(\mf x) \hat{a}^{\tc{b}\dagger}_1.
\end{align}
We then see that the final state of the fields $\tc{A}$ and $\tc{B}$ can be written as
\begin{align}
    \hat\rho_\tc{d} = \tr_{\phi,H}(\hat\rho_f) = \hat{\rho}_{\tc{d},0} + \tr_\phi\left(\hat{U}_I^{(1)} \hat{\rho}_{\tc{d},0}\hat{U}_I^{(1)^\dagger} + \hat{U}_I^{(2)} \hat{\rho}_{\tc{d},0}+ \hat{\rho}_{\tc{d},0}\hat{U}_I^{(2)^\dagger}\right) + \mathcal{O}(\lambda^4),
\end{align}
where
\begin{align}
    \hat{U}_I^{(1)} &= -\ii \int \dd V \, \hat{h}_\text{eff}(\mf x),\\
    \hat{U}_I^{(2)} &= -\int \dd V \dd V' \, \hat{h}_\text{eff}(\mf x)\hat{h}_\text{eff}(\mf x') \theta(t-t'),
\end{align}
with
\begin{equation}
    \hat{h}_\text{eff}(\mf x) = \lambda \hat{Q}_1^{\tc{a}}(\mf x) \hat{\phi}(\mf x) +\lambda \hat{Q}_1^{\tc{b}}(\mf x) \hat{\phi}(\mf x)  = \lambda(\Lambda_\tc{a}(\mf x) \hat{a}^{\tc{a}}_1+\Lambda^*_\tc{a}(\mf x) \hat{a}^{\tc{a}\dagger}_1 + \Lambda_\tc{b}(\mf x) \hat{a}^{\tc{b}}_1+\Lambda^*_\tc{b}(\mf x) \hat{a}^{\tc{b}\dagger}_1)\hat{\phi}(\mf x).
\end{equation}
This is exactly the leading order result when one considers the interaction of two harmonic oscillators interacting with a quantum field $\hat{\phi}(\mf x)$. That is, the final state of the modes can be written as
\begin{equation}
    \hat{\rho}_\tc{d} = \tr_\phi(\hat{U}_I (\hat{\rho}_{\tc{d},0} \otimes \hat{\rho}_\phi )\hat{U}_I^\dagger) + \mathcal{O}(\lambda^4), \quad \hat{U}_I = \mathcal{T}\exp\left(-\ii \int \dd V \, \hat{h}_\text{eff}(\mf x)\right).
\end{equation}
The leading order computations can then be carried on analogously to harmonic oscillator particle detectors (for details on this calculation, see e.g.~\cite{EricksonZero}). 


\section{Estimating levels of mixedness}\label{app:mixed}
An important objection was raised in~\cite{max} against the possibility of harvesting at weak coupling with local modes of a relativistic field theory. This objection is ultimately a consequence of the fact that, due to the Reeh-Schlieder property, the reduced state of a relativistic field on any single mode in a compactly supported region is mixed, and having the probes initialized in a mixed state is known to hinder entanglement harvesting~\cite{maxReply}. However, the presence of a potential that effectively traps the field in a localized region of space can drastically reduce the initial mixedness in the probe, thus bringing the localized mode closer to its ground state. In particular, if we have access to a localized quantum field and we use one of the normal modes of the field as the detector observable, the objection disappears altogether since the initial state of the probe is pure by construction if the global state for the field is the vacuum. 

More generally, in order to address these claims quantitatively in more detail, we can evaluate how fast the mixedness of a given localized mode varies with the size of the mode's support, at a fixed strength of the potential. Let us consider a probe field $\phi_{\tc{d}}$ in Minkowski space, and take the field mode defined at an instant of time $t=0$ by the quadratures
\begin{align}
    \hat Q &= \int \dd^d \bm{x}\, \hat\phi_{\tc{d}}(0, \bm{x})f(\bm{x}), \\
    \hat P &= \int \dd^d \bm{x}\, \hat \pi_{\tc{d}}(0, \bm{x})f(\bm{x})
\end{align}
where $\hat \pi_{\tc{d}}(t, \bm{x}) = \partial_t \hat \phi_{\tc{d}}(t, \bm{x})$ is the conjugate momentum to $\hat\phi_{\tc{d}}$, and $f(\bm{x})$ is a strongly localized function satisfying
\begin{equation}
    \int \dd^d \bm{x}\,f(\bm{x})^2 = 1.
\end{equation}
The $L^2(\mathbb{R}^d)$ normalization is of course chosen in order to guarantee $[\hat Q, \hat P] = \ii\mathds{1}$. The mixedness of this mode is fully characterized by the symplectic eigenvalue of its covariance matrix---which is simply the covariance matrix of the full probe field reduced to this chosen degree of freedom. Since there are no correlations between position and momentum of the field, this symplectic eigenvalue acquires a simple expression,
\begin{equation}\label{eq:sympval}
    \nu = 2\sqrt{\expval{\hat{Q}^2}\expval{\hat{P}^2}}
\end{equation}
where the expectation values above are taken with respect to the global vacuum state of the probe field. By expanding the probe field on its basis of normal modes, we can write 
\begin{align}
    \hat{Q} &= \sum_{\bm n} c_{\bm n} \hat{\phi}_{\bm n}, \\
    \hat{P} &= \sum_{\bm n} c_{\bm n} \hat{\pi}_{\bm n}
\end{align}
where
\begin{equation}
    c_{\bm n} = \int \dd^d \bm{x}\,f(\bm{x})v_{\bm n}(\bm{x}),
\end{equation}
and $v_{\bm{n}}(\bm x)$ form a basis of real functions built from the spatial profiles $\Phi_{\bm{n}}(\bm x)$, that can be chosen to be orthonormal. The normalization condition for $f(\bm{x})$ and the fact that the basis $\{v_{\bm n}(\bm{x})\}$ is orthonormal then implies that
\begin{equation}
    \sum_{\bm n} c_{\bm n}^2 = 1.
\end{equation}
The global state of the field is just the tensor product of the ground states of each of the basis modes
\begin{equation}
    \ket{0_{\tc{d}}} = \bigotimes_{\bm n} \ket{0_{\bm n}},
\end{equation}
and we know that for each mode we have
\begin{align}
    \expval{\hat \phi_{\bm n}^2} &= \dfrac{1}{2\omega_{\bm n}}, \\
    \expval{\hat \pi_{\bm n}^2} &= \dfrac{\omega_{\bm n}}{2}.
\end{align}
Therefore, the expectation values $\langle\hat Q^2\rangle$ and $\langle\hat P^2\rangle$ can be very simply written as
\begin{align}
    \langle\hat Q^2\rangle &= \dfrac{1}{2}\sum_{\bm n} \dfrac{c_{\bm n}^2}{\omega_{\bm n}}, \\
    \langle\hat P^2\rangle &= \dfrac{1}{2}\sum_{\bm n} \omega_{\bm n} c_{\bm n}^2.
\end{align}
By plugging this into~\eqref{eq:sympval}, we obtain
\begin{equation}
    \nu = \left[\left(\sum_{\bm n} \dfrac{c_{\bm n}^2}{\omega_{\bm n}}\right)\left(\sum_{\bm m} \omega_{\bm m}c_{\bm m}^2\right)\right]^{1/2}.
\end{equation}
The expression above makes it clear that $\nu$ satisfies $\nu\geq 1$ (as it must) thanks to the Cauchy-Schwarz inequality,
\begin{equation}
    ||u||\cdot||v|| \geq |\langle u, v\rangle|
\end{equation}
where we take $u$ and $v$ to be vectors with components $u_{\bm n} = c_{\bm n}/\sqrt{\omega_{\bm n}}$ and $v_{\bm n} = \sqrt{\omega_{\bm n}}c_{\bm n}$ respectively, $\langle u, v\rangle$ is the standard Euclidean inner product, and we note that
\begin{equation}
    \langle u, v\rangle = \sum_{\bm n} c_{\bm n}^2 = 1.
\end{equation}
 Furthermore, by recalling that the Cauchy-Schwarz inequality is saturated only when $u$ and $v$ are proportional to each other, we see that $\nu = 1$ (i.e., the chosen mode of the field is pure) only when its spatial profile $f(\bm{x})$ has nonzero overlap exclusively with basis modes of the same eigenfrequency. If $f(\bm{x})$ includes eigenfunctions associated with more than one frequency, the reduced state of the mode $(\hat Q, \hat P)$ is mixed. 
 
 For a free field in Minkowski space, one can choose the basis of normal modes of the field to consist of plane waves (or sines/cosines if we want so select real spatial profiles), in which case $c_{\bm n}$ is directly related to the Fourier transform of $f(\bm{x})$. Since any compactly supported function in position space cannot be compactly supported in Fourier space, it follows that any localized spatial profile will necessarily include nonzero coefficients of arbitrarily high frequencies---which, according to the last comment made on the previous paragraph, means that the symplectic eigenvalue for the reduced state of the local mode $(\hat Q, \hat P)$ must be strictly larger than $1$. This is a very simple particular case of the general fact stated in~\cite{max} that any mode of a free field theory that only has nonzero support over a finite region of space is necessarily mixed\footnote{Note that, even considering this observation, the impact on entanglement harvesting is less than one might be led to believe at first. As noted in~\cite{maxReply}, even in flat space, it is possible to build a sequence of localized field modes whose symplectic eigenvalues can be made arbitrarily close to $1$. Therefore, although perfect purity is never attained in a strict sense, there is no nontrivial lower bound to the purity of a local mode, even if its spatial profile is compactly supported.}. 
 
\begin{figure}[h!]
        \centering
        \includegraphics[width=8.6cm]{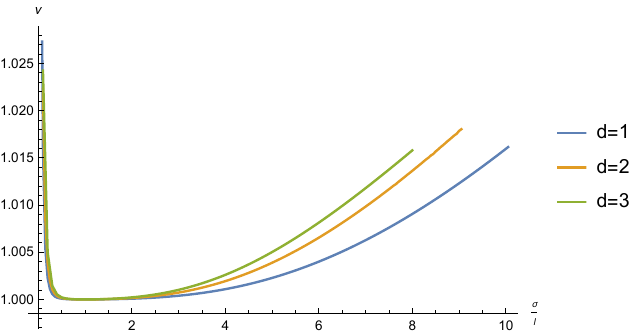}
        \caption{Symplectic eigenvalue of the mode defined with the spatial profile~\eqref{eq:gaussiansmearing} in $1$, $2$, and $3$ dimensions, as a function of the ratio $\sigma/\ell$. For concreteness, the mass of the field was set so that $m\ell = 10$ in all cases.}
        \label{fig:sympeigenvalues}
    \end{figure}

 By adiabatically turning on an external confining potential, however, one can systematically take the localized mode closer to a true normal mode of the field, and thus effectively bring it closer to its ground state. For concreteness, let us now consider a localized mode whose spatial profile is given by
\begin{equation}\label{eq:gaussiansmearing}
    f(\bm{x}) = \dfrac{1}{\pi^{d/4}\sigma^{d/2}}e^{-|\bm x|^2/2\sigma^2},
\end{equation}
where $\sigma$ is a quantity with dimensions of length which determines the effective size of the spatial profile of the mode (i.e., the radius of the region where the mode is strongly supported). The probe field will be placed in an external potential $V(\bm{x})$ given by
\begin{equation}\label{eq:harmonicpotential}
    V(\bm{x}) = \dfrac{|\bm{x}|^2}{2\ell^4}.
\end{equation}
The spatial profile for lowest-frequency mode of the field in this case is simply given by
\begin{equation}
    v_0(\bm{x}) = \dfrac{1}{\pi^{d/4}\ell^{d/2}}e^{-|\bm x|^2/2\ell^2}.
\end{equation}
Now, it should be clear that the spatial profile $f(\bm{x})$ simply corresponds to the wavefunction for the ground state of a harmonic oscillator of different natural frequency---i.e., a \emph{squeezed} ground state. From our knowledge of the quantum harmonic oscillator, this immediately allows us to write down what the overlap coefficients $c_{\bm n}$ are. In terms of the quantum numbers in Cartesian coordinates $\bm{n} = (n_1, \dots, n_d) \in \mathbb{N}^d$ in $d$ spatial dimensions, we have that only the all-even ones contribute: 
\begin{equation}\label{eq:overlapfunctions}
    c^2_{2\bm{n}} = \dfrac{1}{(\cosh r)^d}\left(\dfrac{\tanh^2 r}{4}\right)^{\sum_i n_i}\prod_{i=1}^d\left(\dfrac{(2 n_i)!}{(n_i!)^2}\right),
\end{equation}
and the squeezing parameter $r$ is related to the length scales $\sigma$ and $\ell$ by
\begin{equation}
    r = \log\left(\dfrac{\sigma}{\ell}\right).
\end{equation}
Finally, we recall that the normal frequencies of the probe field are determined by
\begin{equation}
    \omega_{\bm{n}} = \sqrt{m^2 + \dfrac{2}{\ell^2}\left(\sum_{i=1}^d n_i + \dfrac{d}{2}\right)}.
\end{equation}
Putting this all together, we can write down the symplectic eigenvalue of the local mode with spatial profile~\eqref{eq:gaussiansmearing} as
\begin{align}
    \nu^2 = \dfrac{1}{(\cosh^2 r)^d}&\left(\sum_{\bm{n} \in \mathbb{N}^{d}}\sqrt{m^2 + \dfrac{2}{\ell^2}\left(2\sum_{i=1}^d n_i + \dfrac{d}{2}\right)}\left(\dfrac{\tanh^2 r}{4}\right)^{\sum_i n_i}\prod_{i=1}^d\left(\dfrac{(2 n_i)!}{(n_i!)^2}\right)\right) \nonumber \\
    \times &\left(\sum_{\bm{n} \in \mathbb{N}^{d}}\dfrac{1}{\sqrt{m^2 + \dfrac{2}{\ell^2}\left(2\sum_{i=1}^d n_i + \dfrac{d}{2}\right)}}\left(\dfrac{\tanh^2 r}{4}\right)^{\sum_i n_i}\prod_{i=1}^d\left(\dfrac{(2 n_i)!}{(n_i!)^2}\right)\right).
\end{align}
By replacing $\sum n_i = n$, we can turn each expectation value into a single sum, and write
\begin{align}\label{eq:sympeigenvaluegaussian}
    \nu^2 = \dfrac{1}{(\cosh^2 r)^d}&\left(\sum_{n=0}^\infty\sqrt{m^2\ell^2 + 4n + d}\left(\dfrac{\tanh^2 r}{4}\right)^{n}F_d(n)\right) \nonumber \\
    \times &\left(\sum_{n=0}^\infty\dfrac{1}{\sqrt{m^2\ell^2 + 4n + d}}\left(\dfrac{\tanh^2 r}{4}\right)^{n}F_d(n)\right),
\end{align}
where we define
\begin{equation}
    F_d(n) = \sum_{\substack{\bm{n}\in \mathbb{N}^d, \\ \sum {n_i} = n}}\prod_{i=1}^d\binom{2n_i}{n_i}.
\end{equation}    
The series can then be evaluated numerically, with the result in $1$, $2$, and $3$ dimensions being displayed in Fig.~\ref{fig:sympeigenvalues}. We see that, for a fixed value of the ratio $\sigma/\ell$, the mode deviates from purity faster in higher dimensions, but for a reasonably large interval---with $\sigma$ and $\ell$ differing by almost a factor of $10$---the symplectic eigenvalue $\nu$ stays within $5\%$ of perfect purity.

\twocolumngrid

\bibliography{references.bib}

\end{document}